\documentclass{elsart}
\usepackage{natbib}

\usepackage{epsfig}
\usepackage{multirow}
\usepackage[figuresright]{rotating}

\def\SoneA{\mbox{\bf S1A}}
\def\SoneB{\mbox{\bf S1B}}
\def\Stwo{\mbox{\bf S2}}
\def\tof{\mbox{\bf TOF}}
\def\Geant{\mbox{GEANT}}
\def\G4{\mbox{GEANT 4}}
\def\Tina{\mbox{TINA}}

\begin{document}
\runauthor{bham, bnl and cern }
\begin{frontmatter}
\title{The scattering of muons in low Z materials.}
\author[bham]{D.Attwood}
\author[bham]{P.Bell\thanksref{now-man}}
\author[bham]{S.Bull}
\author[bham]{T.McMahon}
\author[bham]{J.Wilson}
\author[bnl]{R.Fernow}
\author[cern]{P.Gruber\thanksref{now-stg}}
\author[icstm]{A.Jamdagni}
\author[icstm]{K.Long}
\author[icstm]{E.McKigney\thanksref{now-lanl}}
\author[icstm]{P.Savage}
\author[ral]{M.Curtis-Rouse}
\author[ral]{T.R.Edgecock}
\author[ral]{M.Ellis\thanksref{now-fnal}}
\author[ral]{J.Lidbury}
\author[ral]{W.Murray}
\author[ral]{P.Norton}
\author[ral]{K.Peach\thanksref{now-adams}}
\author[riken]{K.Ishida}
\author[riken]{Y.Matsuda}
\author[riken]{K.Nagamine}
\author[riken]{S.Nakamura}
\author[triumf]{G.M.Marshall}
\author[ucla]{S.Benveniste}
\author[ucla]{D.Cline}
\author[ucla]{Y.Fukui}
\author[ucla]{K.Lee}
\author[ucla]{Y.Pischalnikov}
\author[oxford]{S.Holmes}
\author[tjnaf]{A.Bogacz}

\address[bham]{University of Birmingham}
\address[bnl]{Brookhaven National Laboratory}
\address[cern]{CERN}
\address[icstm]{Imperial College of Science Technology and Medicine}
\address[ral]{CCLRC Rutherford Appleton Laboratory}
\address[riken]{Riken}
\address[triumf]{TRIUMF}
\address[ucla]{University of California at Los Angeles}
\address[oxford]{Oxford University}
\address[tjnaf]{Jefferson Laboratory}
\thanks[now-man]{Now at the University of Manchester}
\thanks[now-stg]{Now at University of St. Gallen}
\thanks[now-lanl]{Now at Los Alamos National Laboratory}
\thanks[now-fnal]{Now at Fermi National Accelerator Laboratory}
\thanks[now-adams]{Now at the Adams Institute}

\begin{abstract}
\noindent
  This paper presents the measurement of the scattering of 172~MeV/c muons 
in assorted materials, including liquid hydrogen, motivated by the need to 
understand ionisation cooling for muon acceleration.

  Data are compared with predictions from the \G4\ simulation code and 
this simulation is used to deconvolute detector effects. The scattering 
distributions obtained are compared with the Moliere theory of multiple 
scattering and, in the case of liquid hydrogen, with ELMS. With the 
exception of ELMS, none of the models are found to provide a good 
description of the data. The results suggest that ionisation cooling will 
work better than would be predicted by \Geant~4.7.0p01.


\end{abstract}
\begin{keyword}
multiple scattering; ionisation cooling
\end{keyword}
\end{frontmatter}






\section{Introduction}

The storage of intense muon beams is required for the construction of a
Neutrino Factory or muon collider.
The ionisation cooling technique\cite{ref:cooling} is a leading contender to
maximise the intensities for 
the Neutrino Factory and it is even more likely to be used at a muon collider.
This technique relies upon the cooling effect of dE/dx losses in
low Z materials coupled with R.F. acceleration to reduce the emittance of
 a beam. Multiple scattering in the material  heats the beam 
and the overall cooling is thereby reduced.
It is  therefore  important to confirm the theoretical
understanding and computational implementations of multiple scattering
of muons of around 200~MeV/c.
There is no published measurement of muon scattering in this region,
and
furthermore   data on electron scattering with similar
$\beta$\cite{ref:andrievsky} suggest an excess  in the tails compared
with predictions from Moliere theory\cite{ref:moliere}, especially
for low $Z$ materials. This was first observed in ref.~\cite{ref:fernow}.

Many problems involving  multiple scattering have employed \Geant~4
\cite{ref:g4} to evaluate the effects.
A new model of scattering, ELMS\cite{ref:elms},
 has recently become available. This is a return to electromagnetic
first principles, but often uses numerical integration rather than closed-form 
approximate integrals. 

The MuScat experiment was accepted by TRIUMF in 1999 to run in the
 M11 muon beam-line,
and an engineering run took place in summer 2000. 
A new scintillating
fibre tracker was built for the physics run. The data presented here 
were collected  in April and May  2003, in the M20 muon beam-line in TRIUMF.

Section~\ref{sec:hardware} contains a description of the principle and 
technology  of the MuScat experiment. This is followed in sections
~\ref{sec:scifi}, \ref{sec:calorimeter} and \ref{sec:tof} 
by details of the performance of the tracker, calorimeter and scintillator
 subsystems.
The beam properties are described in section~\ref{sec:beam}.
These descriptions are accompanied by  discussion of the \G4\ model
 which was used to deconvolute the detector effects, a process
discussed in section~\ref{sec:analysis}.
The data are presented in section~\ref{sec:results}, and  conclusions
can be found in
section~\ref{sec:conclusions}.

\begin{figure}
\begin{center}
\begin{picture}(390,230)
\put(3,-40){\mbox{
\epsfig{figure=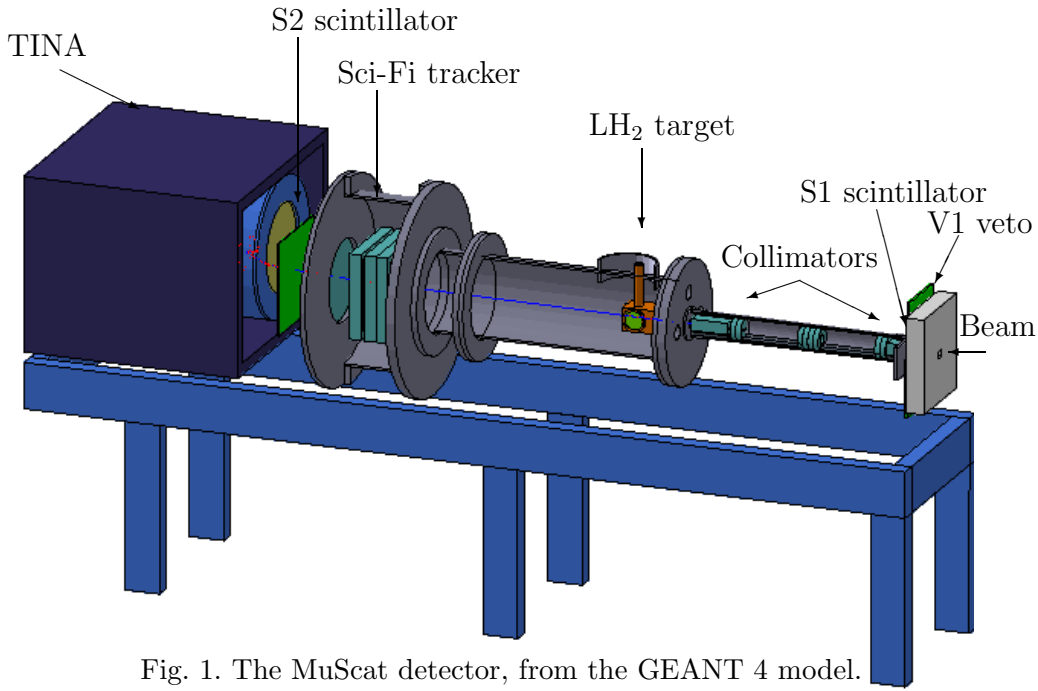,width=13.0cm}
}}
\put(10,200){\mbox{\Tina}}
\put(30,195){\vector(2,-1){30}}
\put(110,210){\mbox{S2 scintillator}}
\put(120,207){\vector(0,-1){62}}
\put(135,190){\mbox{Sci-Fi tracker}}
\put(150,187){\vector(0,-1){40}}
\put(230,170){\mbox{LH$_2$ target}}
\put(250,165){\vector(0,-1){30}}
\put(280,120){\mbox{Collimators}}
\put(310,115){\vector(-3,-1){20}}
\put(310,115){\vector(2,-1){30}}
\put(310,145){\mbox{S1 scintillator}}
\put(339,142){\vector(1,-4){11}}
\put(358,133){\mbox{V1 veto}}
\put(370,132){\vector(-1,-2){9}}
\put(370,94){\mbox{Beam}}
\put(380,88){\vector(-4,0){14}}
\end{picture}
\vspace*{0.5cm}
\end{center}
\caption{\label{fig:overview}The MuScat detector, from the \G4\ model.}
\end{figure}

The detector performance has been modelled using \Geant~4.7.0p01.
 This toolkit
was used to make a complete description of the geometry shown in
figure~\ref{fig:overview}.
Considerable attention has been attached to the  simulation of the 
collimator system and the fibre tracker, which are critical to the experiment
and whose performances are sufficiently complex that they are hard to 
extract from the data alone.

\section{The Experimental Apparatus}
\label{sec:hardware}

The technique of the MuScat experiment was to prepare a narrow collinear
muon beam by collimation, and to use it to illuminate a variety
of target materials. 
The position of the outgoing particles was measured
after travelling
a sufficient distance that the position and angle were highly correlated.

The key components are the collimation system, a wheel with a selection of
solid targets and an optional liquid hydrogen (LH$_2$) vessel, a
scintillating fibre tracker, and the trigger scintillators. In addition to
these the \Tina\ (TRIUMF Iodide of Sodium) calorimeter\cite{ref:tina}
 was useful to help understand the system.
There were also some extra scintillators employed as veto counters.

The  passive components have been
 implemented in a \G4\ model in a simplified manner, sufficient
to simulate the
materials which would affect the observed scattering distributions. 
The
 position of many externally visible components were checked with a precision
of a few mm in the M20 beam-line in TRIUMF, and no deviations found.
The position of the  target wheel is  known from the engineering drawings.
The locations of the scintillators and the \Tina\ calorimeter are 
known from the measurements.

The  coordinate system chosen is that z is defined approximately
by the beam direction, y is vertical and x is horizontal so 
 as to define a right handed coordinate system. Since the collimator slits are 
narrow in the vertical direction, the y coordinate of a muon is the 
measure of its scattering. The precise  definition of the $z$ axis 
is that that the first two detector planes are both centred on 
$x =  y = 0$.

\subsection{The M20 beam-line}

The experiment used the M20 beam-line at TRIUMF. The extracted proton
beam from the cyclotron interacted in a target and produced pions at a
production angle of 55 degrees. The beam-line had a muon decay channel 
consisting of two dipoles with
quadrupole focusing. The beam-line was tuned to optimise the capture of
high momentum muons from forward pion decays. The mean beam
momentum was 172$\pm$2.0~MeV/c and the accepted momentum bite was
approximately 1~MeV/c.
The beam delivered to the experiment consisted of muons with a 
small pion background, but a mixed beam could be produced, as described later
in the discussion of the time of flight system. The trigger rate
was 80 particles/sec; the particle rate incident on the first scintillator,
before collimation,
was three orders of magnitude larger. The spot size at the
first collimator was 50~mm in diameter.


\subsection{The Collimator}
\label{sec:coll}

\begin{figure}[htp]
\begin{center}
\epsfig{figure=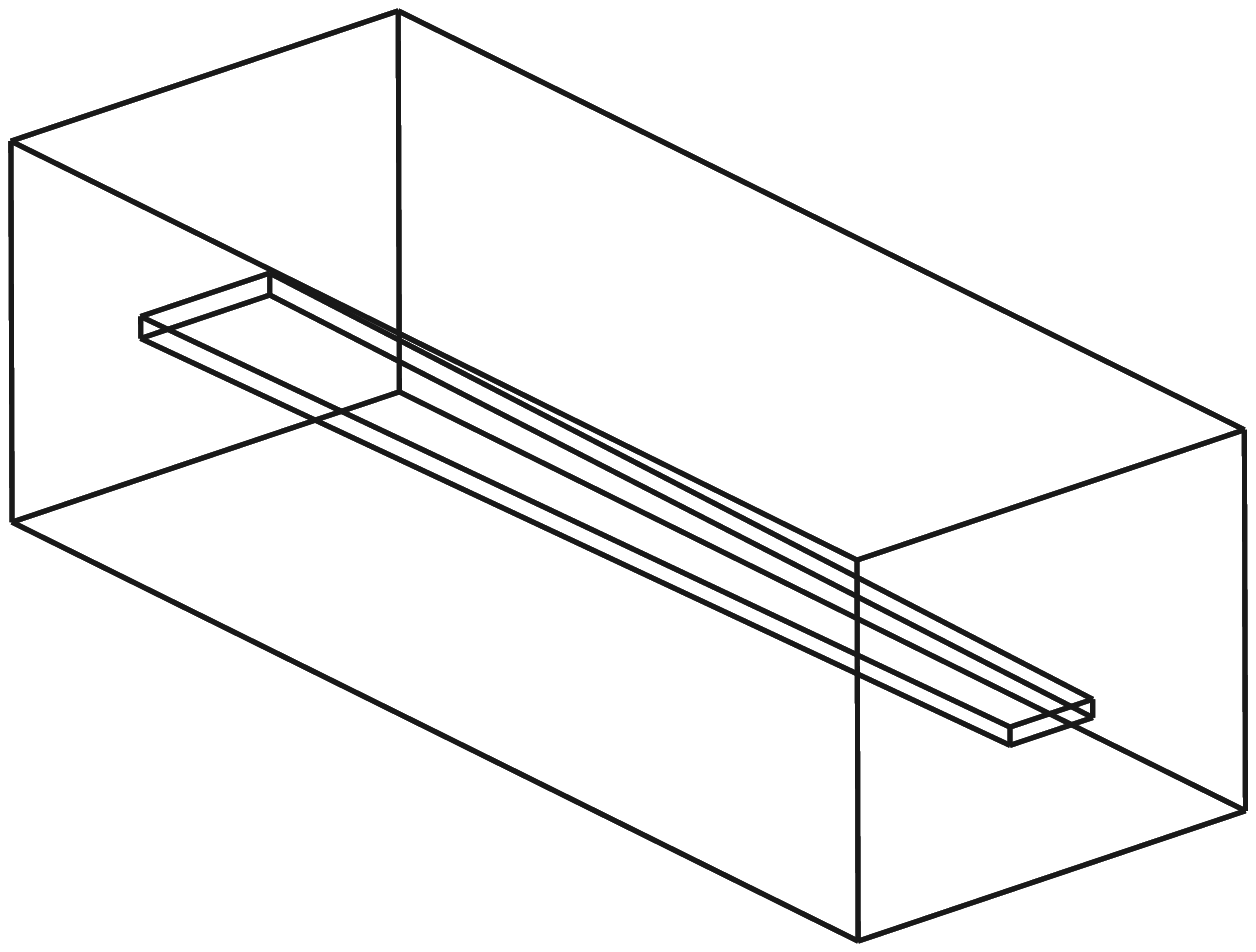,width=6.8cm}
\epsfig{figure=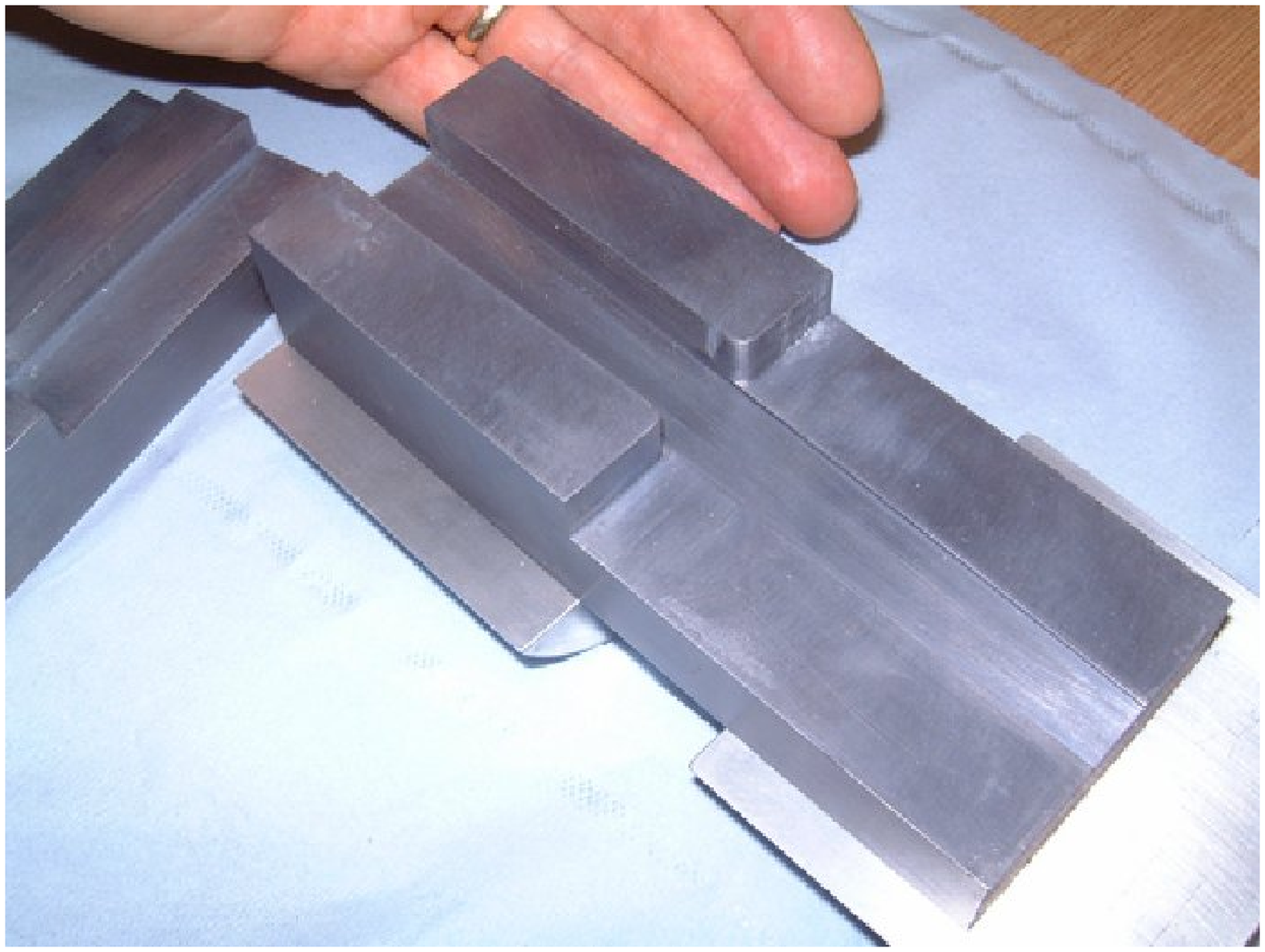,width=6.8cm}
\end{center}
\caption{\label{fig:collimator2} The second collimator piece, made
of  160~mm of lead. On the  left is the computer model,
and on the right  the two interlocking pieces of lead used to
construct it.}
\end{figure}

The collimator system consisted of  two lead blocks with slits in them,
augmented by other active and passive collimators.
The
accepted particles had to pass through both slits. This defined a beam which
was
narrow in one direction, vertically, but much broader in the other to maintain 
rate.
The upstream lead block was 40~mm thick, with a  simple slit, 20~mm by 2~mm.
The second block, shown in figure~\ref{fig:collimator2},
 had a trapezoidal slit, 12.8~mm by 2.0~mm at the upstream end,
expanding
to 20.0~mm by 2.88~mm at the downstream end.
The collimator blocks were each made from
two interlocking pieces of lead, with the lower one fixed to two steel
rods by aluminium supports. The distance between the outer faces of
the collimators was 963~mm.

The system was behind a shield wall; 8~cm thick steel with a 4 by 3~cm 
opening aligned with the lead collimators.
There were also eight auxiliary collimators, made from lead discs 20~mm thick
and with 20~mm diameter openings,
which were mounted on the steel rods between the two main collimators.
They were intended to intercept any particle which might have scattered
in the jaws of the first collimator and then missed the second one entirely.
They were designed so particles scattered from them could not pass 
through the slit in the final collimator without another interaction.
The entire tube was wrapped in lead sheet to a thickness of 6~mm.
This ensures that all muons reaching the tracker have passed through the
entire collimator system.

Finally, the downstream passive lead collimator was supplemented with two
scintillators, 3~mm by 3~mm by 30~mm. These were positioned just at
the front face of the collimator, above and below the slit.
They were designed to record particles which might otherwise have hit the
lead and then emerged into the slit with a large scattering and energy loss.
They were read-out by 7 of the clear 1~mm fibres used for the Sci-Fi tracker.
The fibres were attached to each end and  taken to a photomultiplier 
tube (PMT).

\begin{figure}[htp]
\begin{center}
\epsfig{figure=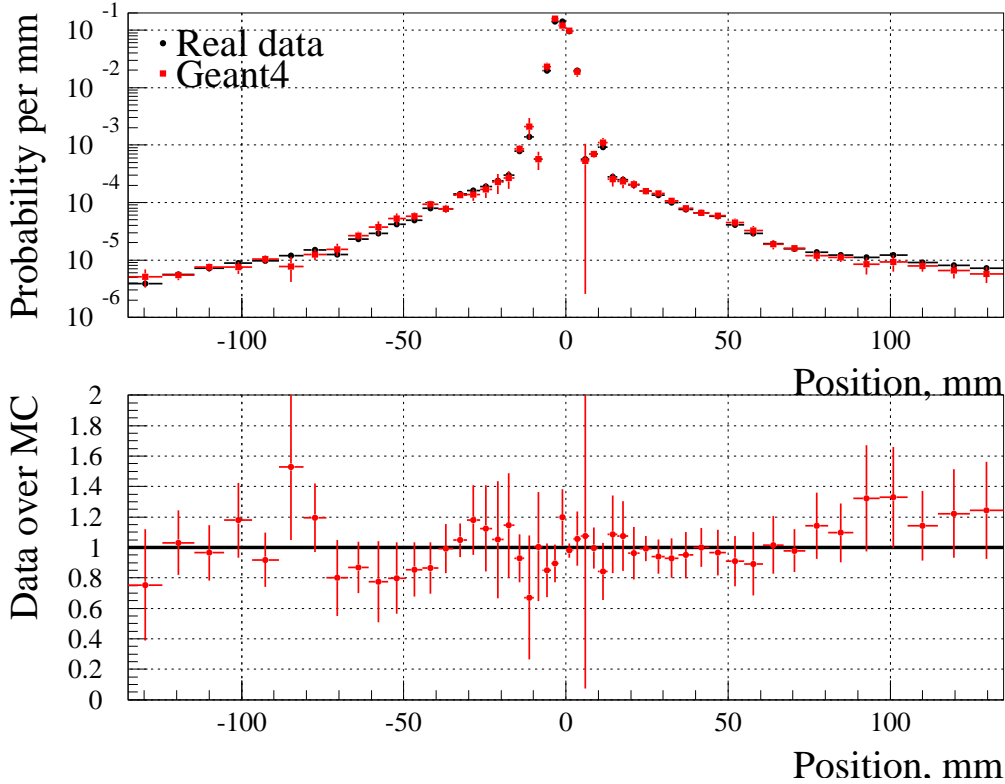,width=13.8cm}
\end{center}
\caption{\label{fig:bare_collimator} The distribution of particles at detector
plane 1 in data and simulation for no target. The error bars include
systematic error estimates. The large error bar just to the right of the peak
includes the effect of possible misalignment.
}
\end{figure}

The width of the collimated beam must be well simulated, and unfortunately
this is not the case with the assumption of a perfect geometry. The
apparent slit width and also the observed tails are both wider in data
than in the simulation. The only credible way to increase the collimated beam
width is to hypothesise that the second collimator was not perfectly closed.
It was constructed as two blocks of lead which dovetail neatly and firmly but
are not attached together. The lower block was held firmly to the rails,
 and the whole assembly is prevented from rotating by a grub screw
through the collimator tube attached to the top lead block. If the
system was somehow pressured after or during installation the grub screw
could pull the top lead back, widening the jaw.
The simulated distribution of particles through the expanded collimator 
assumption is shown in figure~\ref{fig:bare_collimator}, and
this is used hereafter.

This is unfortunate, as not only does the central distribution widen but
the inner faces of the collimator can become directly exposed to incident 
particles, allowing scattering from them. These tails are indeed observed.
The energy distribution of the tail particles, as measured in \Tina,
is consistent with what would be expected if this is the correct explanation.
The size of the movement is  small.
The best representation of the observed data is
obtained if the jaw is opened from 2.0~mm to 2.48~mm.

\subsection{The target system}

There were two separate target systems: a wheel with a choice of solid targets
and a vessel for liquid hydrogen.
The target wheel  was controlled by  a
stepper motor, and  presented
a choice of twelve target positions to the beam.
The position was monitored by an
optical
LED system which allowed readout of the target number and a confirmation that
the  position was exactly correct.
Table~\ref{ta:targets} shows the targets used, and the number of events
selected for analysis from each. 
The density and radiation length, X$_0$, are also shown.

\begin{table}[htbp]
\begin{center}

\begin{tabular}{cccccr}     \hline
No. & Material & Thickness, mm  & Density, g/cc & X$_0$, \% & Events \\ 
\hline \hline 
0   & Lithium   & 12.78          & 0.53    & 0.82  & 805239     \\
1   & Lithium   &  6.43          & 0.53    & 0.41  & 1224756     \\
2   & Lithium   &  6.40          & 0.53    & 0.41  & 882449     \\
3   & Lithium   & 12.72          & 0.53    & 0.81  & 1215336     \\
4   & Beryllium &  0.98          & 1.85    & 0.28  & 500766     \\
5   & Beryllium &  3.73          & 1.85    & 1.06  & 1186528     \\
6   & CH$_2$    &  4.74          & 0.93    & 0.99  & 802426     \\
7   & Carbon    &  2.50          & 1.69    & 1.53  & 801899     \\
8   & Aluminium &  1.50          & 2.70    & 1.69  & 1201280     \\
9   & None      & -              & -       & -     & 2259476     \\
10  & Iron      &  0.24          & 7.86    & 0.82  & 845020     \\
11  & Iron      &  5.05          & 7.86    &28.68  & 1225435     \\
\hline
    &Empty H$_2$& 109.0          & -       & -     & 2694511     \\
    & H$_2$     & 109.0          & 0.0755  & 1.31  & 2267683     \\
    &Empty H$_2$& 159.0          & -       & -     & 812730     \\
    & H$_2$     & 159.0          & 0.0755  & 1.90  & 1127045     \\
\hline
\end{tabular}
\caption[]{
The targets, and the number of data events selected for each.
}
\label{ta:targets}
\end{center}
\end{table}

The  LH$_2$ vessel was crucial to the experiment, because  
 scattering in liquid hydrogen was the most important goal.
This was designed and built by the cryogenic targets group 
at TRIUMF.
 There is a  contribution
to the scattering from the containing windows, and therefore it
it is essential to minimise and  measure this accurately.
The vessel could be filled with liquid or gaseous hydrogen, 
and the measured distributions with gaseous hydrogen are used to characterise
the windows.
The temperature remained between 15 and 23 K for these cycles, and there
was always a slight overpressure.

The vessel was 150~mm by  100~mm by 100~mm. The 150~mm length was bored with
an 8~cm diameter hole, and an identical hole was made through from one side.
These were capped by four identical windows in the sides and ends,
so that by rotating it a choice of hydrogen thicknesses was presented.
The windows bulged by 4.5~mm, with an estimated error of 0.3~mm,
making the total length 109.0$\pm$0.6~mm or 159.0$\pm$0.6~mm.
They were made from mylar, 5 thousands of an inch thick, wrapped in 6 layers
of super-insulation.

\section{The Sci-Fi tracker }
\label{sec:scifi}

The scintillating fibre (Sci-Fi) tracker 
comprised three
double fibre planes, read out by twenty-four HPK multi-anode PMTs
\cite{ref:PMT}.
The planes were about 1100~mm from the target. The first and second planes 
were mounted together, which means the active elements were displaced by 50~mm.
While for the third plane the displacement was  66.9~mm.

\begin{figure}[htp]
\begin{center}
\epsfig{figure=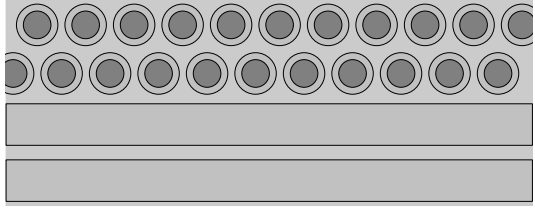,width=7.0cm}
\end{center}
\caption{\label{fig:fibres} The layout of the scintillating fibres. Each plane
consists of both horizontal and vertical fibres, overlapping in two rows 
to give complete coverage. In each row the distance between fibre centres was 
1.17~mm and the distance between the rows was 1.17~mm.
They are potted in black epoxy.
}
\end{figure}

The double planes consisted of 
two planes (each constructed as shown in figure~\ref{fig:fibres}), one 
comprising 512 horizontal fibres and the other the same number
of vertical fibres.
They have a  pitch of 0.585~mm, covering
300~mm$^2$. The fibre diameter was 1~mm, including 
 the outer 4\% cladding. 
They were arranged in double rows, as in figure~\ref{fig:fibres}.
They were held in a  Delrin frame, 350~mm square with a 300~mm square
central area. 2048 holes were drilled with a CNC machine to define the
locations of the two ends of each fibre. 
The fibres were held under tension and potted
in black epoxy. Finally the ends of the fibres  were cut.

\begin{figure}[htp]
\begin{center}
\epsfig{figure=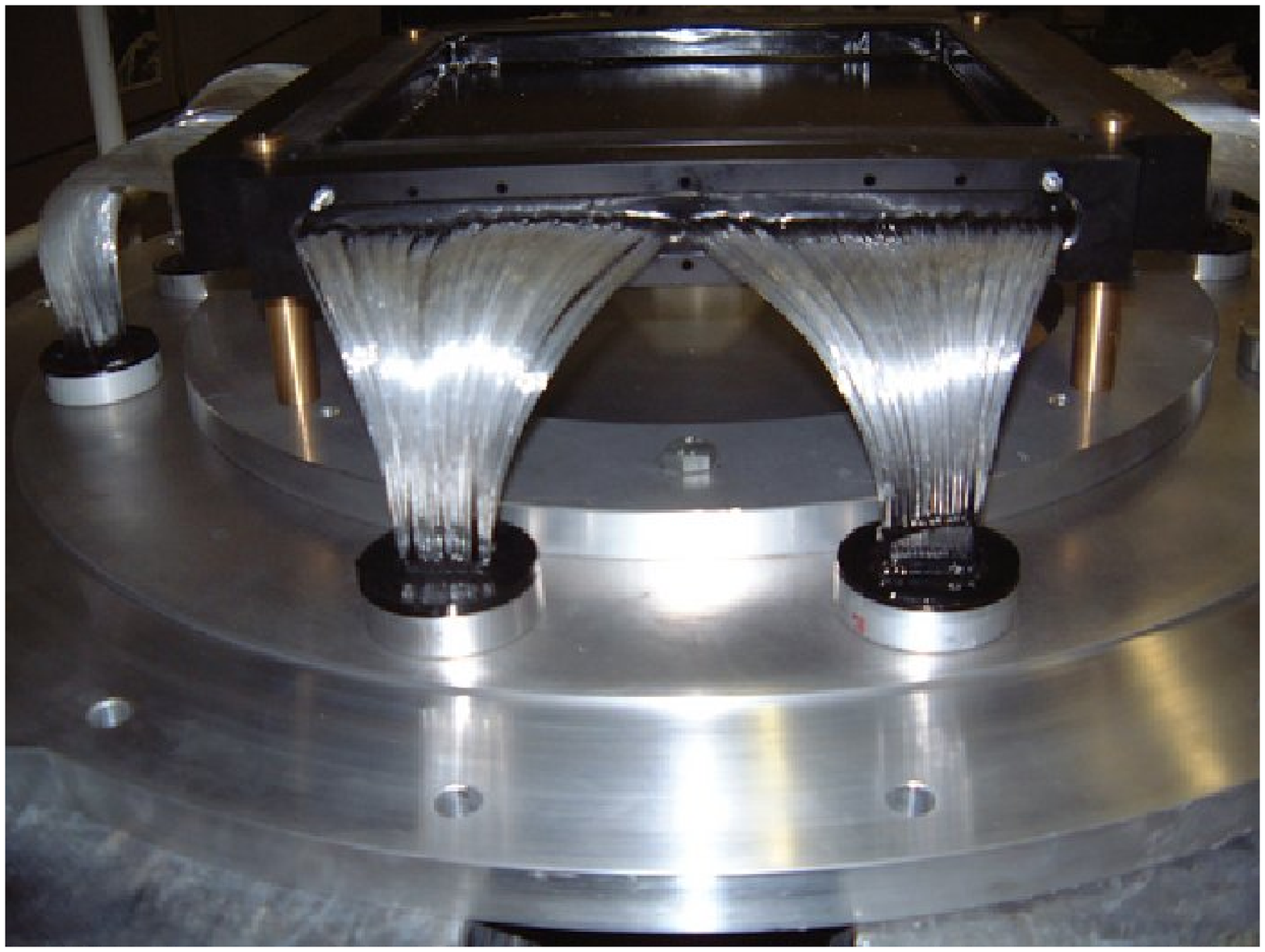,width=6.8cm}
\epsfig{figure=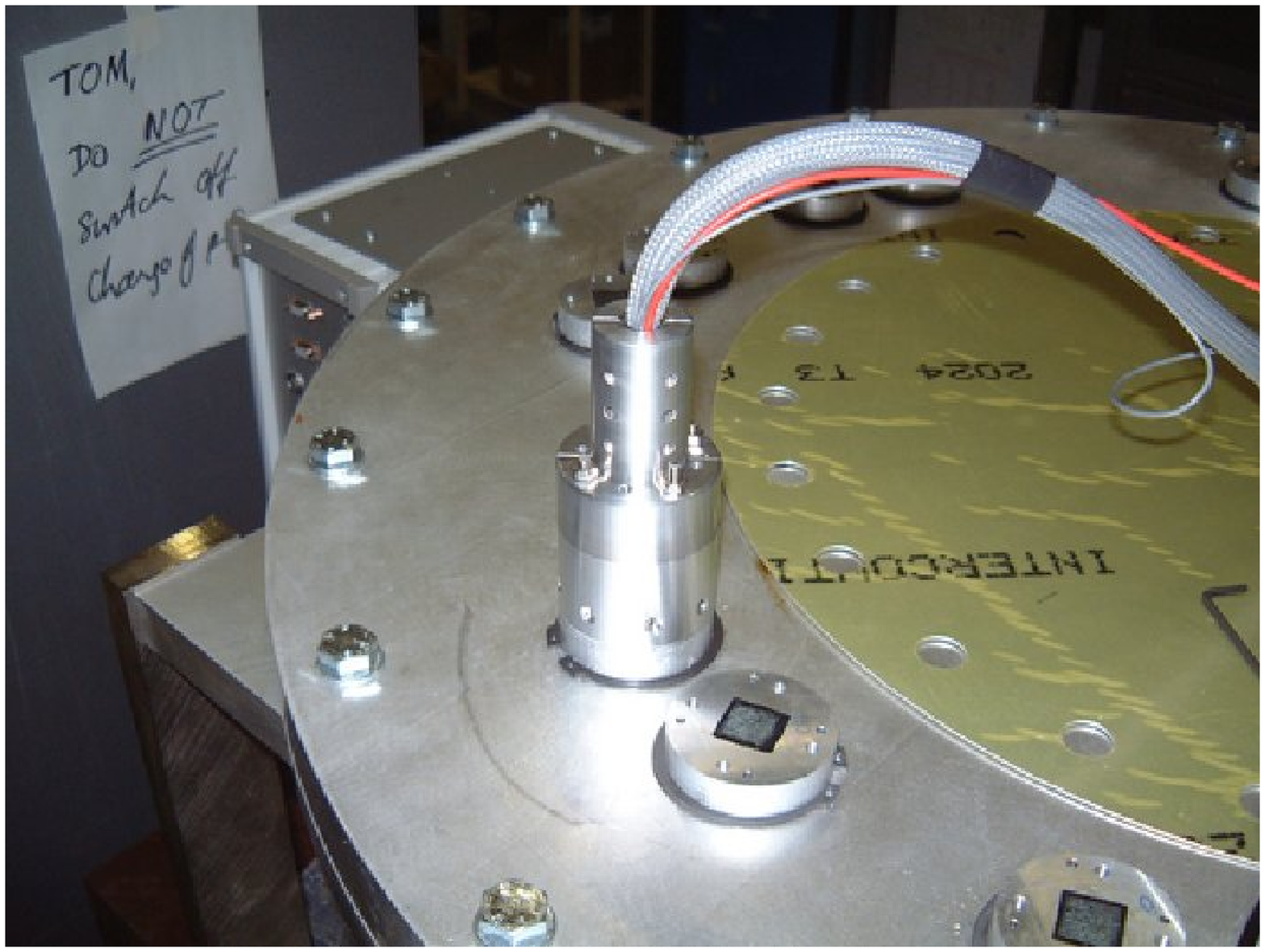,width=6.8cm}
\end{center}
\caption{\label{fig:tracker-pic} The physical arrangement of 
 the Sci-Fi tracker. The left hand figure shows one plane mounted inside
the vessel. The black frame at the top encloses the scintillating fibres,
and the clear routing fibre bundle can be seen. The right hand figure shows
the front faces of the fibre bundles, with a PMT mounted onto one of them.
}
\end{figure}

Each double plane had four Delrin edge connectors, made in a similar manner,
and drilled in one operation with the frame so that the holes matched.
This can be seen in figure~\ref{fig:tracker-pic}~(left).
Into each of these connectors   were potted 512 double-clad 1~mm radius clear
fibres which were  to guide the light to photo-multipliers. 
This was done in two groups of
  256, and the far ends of the fibres, after about 300~mm, were 
arranged into  a square 16 by 16. These were potted in an aluminium
former which was to also act as a vacuum seal for the detector vessel.
The natural tendency of the fibres to close-pack made this construction
difficult, and the edge fibres were not always perfectly placed.
The best bundles were selected for the first detector and for measuring y.
The 16~mm square fibre bundles presented on the outside
of the pressure vessel were connected to the 16-anode PMTs, as shown in
figure~\ref{fig:tracker-pic}~(right).
The readout was done with a custom-made  sample and hold system,
which provided a stable, low-noise readout for the system.
 
Each set of 256 fibres is read out at both ends. One end of the fibres 
is read out by a PMT which measures the signals from 16 adjacent fibres in 
each of its (16) anodes. The other set of ends is distributed across 
the PMT surface so that each PMT anode reads every sixteenth fibre.
In this way, one PMT is used to measure which group of 16 fibres has a
signal and the other which fibre within
the group. This multiplexing scheme can only work with a single charged
particle present.

Simulation of the tracker was fairly complete.
The light yield per fibre in the simulation was tuned to the data 
 using the mean measured 
signal in identified hits using the thick iron target. For a few percent
of fibres with low light yields this does not converge satisfactorily; 
for these the rate of hits is used instead.
Cross-talk and signal shape were  simulated due to
the light spreading in the 1.5~mm glass at the front of the PMT,
the photo-electrons jumping to other dynodes, 
a leaky dynode chain where an electron can skip a stage or move into the
neighbouring chain and
about one percent  of electronic cross-talk.

\begin{figure}[htp]
\begin{center}
\epsfig{figure=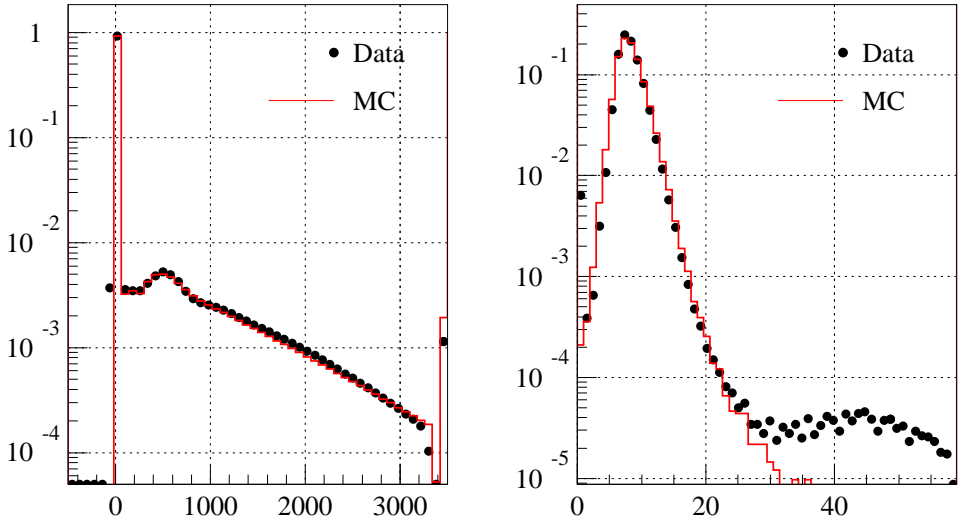,width=13.8cm}
\end{center}
\caption{\label{fig:n_fibres} Left: The number of ADC counts
on every channel in data and simulation with the thick iron target.
The peak at 600 is due to single photo-electrons.
Right: The distribution of the number of
fibres with a signal of over 0.2 photo-electrons observed in data
and simulation.
}
\end{figure}

The distribution  of ADC counts recorded on all channels  
is  in figure~\ref{fig:n_fibres}~(left). The pedestal is clearly visible,
with a peak due to single photo-electrons at around 600 counts which
was used to measure the gain. The upper
edge is cut off by the saturation of the ADCs. The simulation
includes these effects and
shows reasonable agreement.

The distribution of the number of reconstructed fibres with signals in data 
and simulation is shown in figure~\ref{fig:n_fibres} (right).
The data has events with no fibres hit which are not reproduced by simulation.
These are trigger accidental coincidences. There are also events with
several times the normal number, which arise from multiple tracks. Both
types of events are removed in analysis. The central region has a reasonable 
representation.

\begin{figure}[htp]
\begin{center}
\epsfig{figure=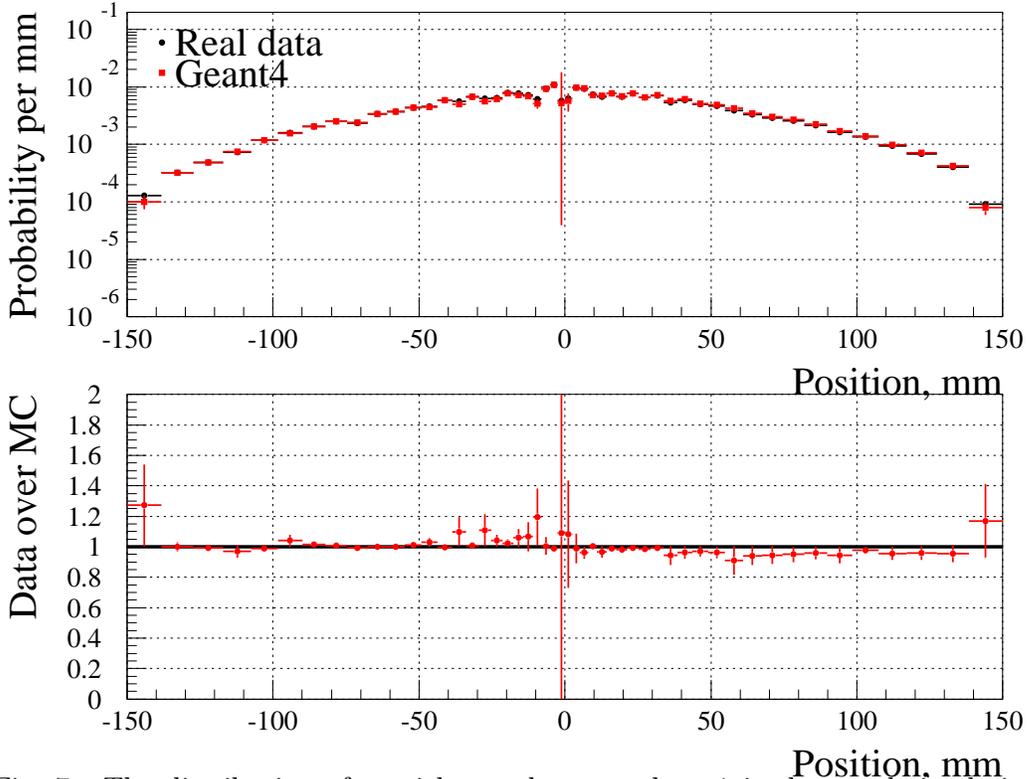,width=13.8cm}
\vspace*{-1cm}
\end{center}
\caption{\label{fig:thick_steel} The distribution of particles at detector
plane 1 in data and simulation when the thick iron target is in place.
}
\end{figure}

The recognition of a hit in the tracker depends upon
the observation of at least one photon at either end, and as the
average light yield is around four photons then any 
poor fibres with lower yields will have significant inefficiency.
This is examined in
figure~\ref{fig:thick_steel}, which shows the
results for a thick iron target, of 28\%~X$_0$.
The scattering distributions in this target are dominated
by the central Gaussian core, and the simulations are therefore
expected to be reliable. The purpose of this target is to illuminate
the entire detector with
tracks for use in alignment and efficiency studies.
There is a distinct drop in efficiency at the very centre of the
distribution, and also some structure at about -10~mm.
The distribution is shown for the entire detector, 
 and the agreement is
 well within 10\% in most bins, with the centre and the 
 edge bins  
showing differences around 20\%. 
It
seems that the simulation of the edges of the PMTs is somewhat defective,
probably due to the difficulty of locating all the clear fibres in
the correct places at the front of the PMT.

\section{The \Tina\ Calorimeter}
\label{sec:calorimeter}

An existing calorimeter from the TRIUMF Laboratory, 
`\Tina'\cite{ref:tina},
was used to understand the beam properties. This is a cylindrical sodium
iodide calorimeter, 460~mm diameter
and 510~mm long, split into two crystals in depth. It is read out using
7 PMTs on the back-face, one in the centre and the other six in a circular
pattern around it. The calorimeter is contained in a metal blockhouse
to reduce external backgrounds. This blockhouse has a small lip that
overlaps the outer edge of the NaI by about 40~mm.
The device  has a measured energy resolution (fwhm) of 3.6\% at 90~MeV for
electrons,
with an energy dependence of $E^{-0.55}$.

As \Tina\ was too small to cover the complete angular acceptance of MuScat,
it was mounted with the centre shifted upwards by 70~mm with respect to
the axis of the experiment. This allowed a complete angular coverage in
the positive y direction. In addition, the energy deposition was integrated
over two time periods: {\tt tina\_early}, corresponding to 500~ns
from the \SoneB\ trigger, and {\tt tina\_total}, corresponding to
10~$\mu$s from the trigger. In general, the first of these will have
only the kinetic energy of the muon, while the second also
contains the electron from the decay of the muon, 
 which has come to rest in TINA

\begin{figure}[th]
\begin{center}
\epsfig{figure=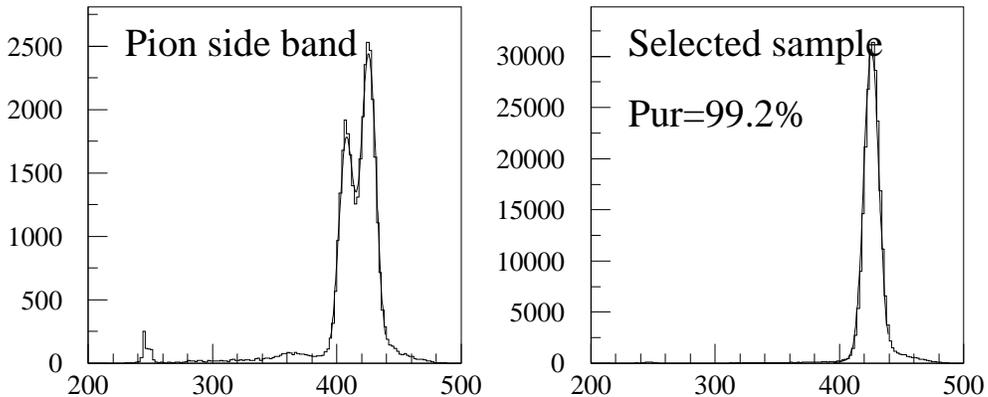,width=13.8cm}
\vspace*{-2cm}
\end{center}
\caption{\label{fig:compos} The {\tt tina\_early} ADC  signal in
the \Tina\ calorimeter for
events selected by timing cuts 
(see section~\ref{sec:composition}) 
to be either mixed pion and muon events
 (left) or the  muon sample used for analysis (right).
A requirement on {\tt tina\_total} - {\tt tina\_early} has been
made so that the muon decay is known to occur after the  {\tt tina\_early}
gate closed, but a small decay tail can still be seen.
 The peak at 245 ADC counts is the pedestal.
}
\end{figure}

The performance of \Tina\ is illustrated by figure~\ref{fig:compos}.
The pedestal is at 245 ADC counts. The left hand plot (pion sideband), 
shows the separation between pions and muon peaks.
Decay electrons have been supressed as discussed in the caption.
If this had not been done there would have been larger tails, and 
an additional 0.4\% would have been assigned to the background.
The \Tina\ calorimeter was modelled in \G4, but 
these simulations   are not used in this analysis.




\section{The Scintillators}
\label{sec:tof}

There were seven instrumented plastic scintillators used in the experiment.
Three of these made up
the trigger system, and the other four were veto
counters.

The trigger was provided by a logical {\tt and} of two overlapped 
finger scintillators, 3~mm square in area and 30~mm long, known
as \SoneA\ and \SoneB. These were  
 mounted  in front
of the collimator to record particles before entering the system.
 They  gave a precise timing signal,
used for the time of flight measurements discussed later.
This was taken in coincidence with the signal from a 330~mm square
scintillator, referred to as \Stwo, which was read out by two PMTs, 
top and bottom, whose results were combined as a logical {\tt or}.
\Stwo\ was placed after the Sci-Fi tracker and before \Tina\ as shown in
figure~\ref{fig:overview}.

Two  large area veto scintillators were butted together just in front of S1;
one had a notch removed that corresponded approximately to S1. These
are referred to as V1. The other two veto scintillators were bars
3~mm by 3~mm by 30~mm which were mounted on the front face of the second 
lead collimator, just above and just below the slot.

The common start for all the TDCs (Time to Digital Converters) 
in the experiment came from the \SoneB\  scintillator. 
A delayed version of \SoneB\  was then read by a TDC. This should have 
given a fixed time, but some events were several ns different and they
were removed by offline selection.
An additional signal from the cyclotron, repeated each 43~ns period, provided
a useful timing reference.

The \Geant\ model included accurate descriptions of the geometry of
the scintillator system. Electronic
handling of the signals, in terms of pulse shaping, and discriminator threshold
for the TDCs was included in an approximate way so that time-walk effects
were reproduced.

In order to reproduce the data, the finger scintillators comprising the
active collimators  had to be 
positioned so that they overhung into the collimator gap slightly:
the top by 190~$\mu$m and the bottom by 70~$\mu$m.
This reinforces the idea that the collimator slot had somehow opened.

\section{Beam properties}
\label{sec:beam}

There has been no attempt to simulate the  beam-line.
However, the x distribution of the observed beam has a very small
dependence upon the initial beam configuration, and for y it is totally 
negligible. The beam divergence is instead determined by
multiple scattering in the \SoneA\ and \SoneB\ 
scintillators, which guarantees a minimum spread, followed by 
the collimation system which limits the maximum angle.

\subsection{Beam momentum determination}

\begin{figure}[th]
\begin{center}
\vspace*{-1.5cm}
\epsfig{figure=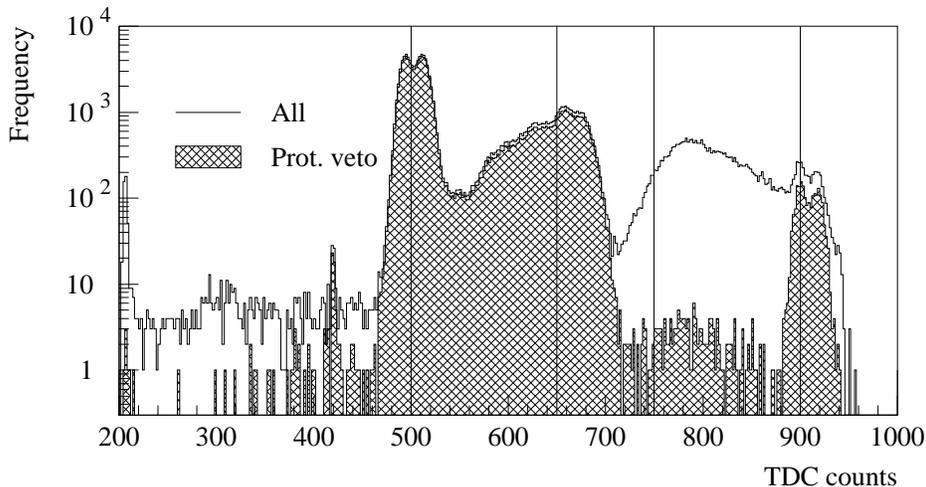,width=13.8cm}
\vspace*{-1.5cm}
\end{center}
\caption{\label{fig:tofmom} The time of arrival of the beam cycle signal
relative to the \SoneB\ scintillator for run 177, which was tuned
to have a mixed particle content. The four lines guide the eye to pions,
muons, protons and electrons. The open histogram is for all events that 
are in time  on \SoneB\, and the
hashed area for events that meet a tighter requirement on \SoneB\, within
0.25~ns of expected, which
excludes protons. A double peak structure can be seen for the precise
peaks; this is not understood.
}
\end{figure}

A special run was taken with the momentum of the second dipole in the beam-line
turned down by 2\%. This meant that the forward muons normally accepted have a
large contamination from other species. The presence of multiple particle
types  has been used
to measure the beam momentum.
 The time of arrival of the cyclotron signal  relative to \SoneB\ is
shown in figure~\ref{fig:tofmom}. 
The  major
peaks are at 500 (pions), 650 (muons), 750 (protons) and 900 (electrons).
In general the fastest particles give the largest times, because what
is measured is the arrival time of the cyclotron signal after  \SoneB.
In the case of protons, the velocity is only 0.2~c, and they arrive seven
beam crossing periods late.
The width in the  electron and pion signals
 are similar, but the others are broader.
For protons this is because the very low velocity leads to a much larger 
spread in time, while the muon arrival time depends upon the decay
point. There is a distinct peak of muons which were produced before
the first bending magnet.

\begin{figure}[th]
\begin{center}
\epsfig{figure=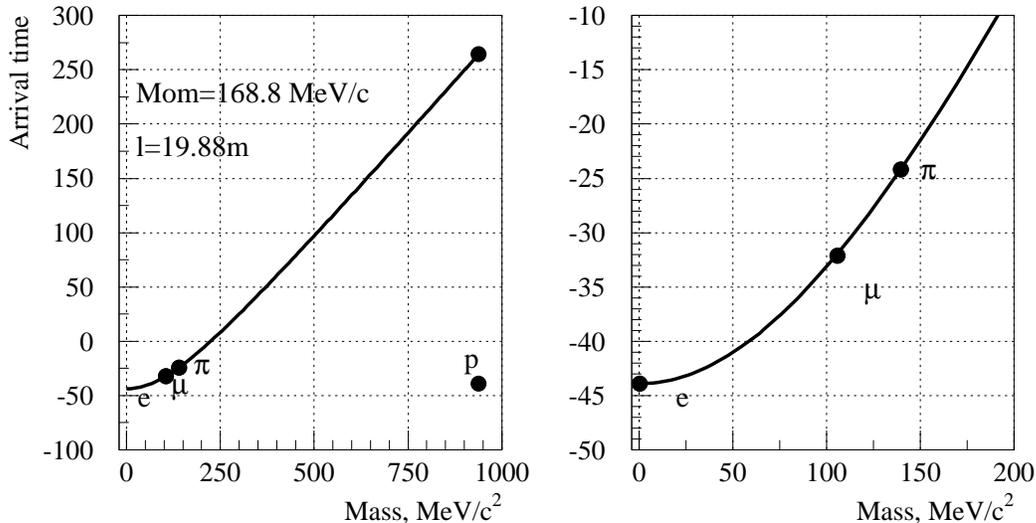,width=13.8cm}
\vspace*{-1cm}
\end{center}
\caption{\label{fig:calcto4}Fitting the beam momentum from the observed peaks
using the cyclotron signal. The negative of the arrival time of the
cyclotron signal is plotted; for protons a point is shown  both at the
measured time and delayed by seven beam cycle periods. }
\end{figure}

The momentum of the beam has been estimated by fitting the arrival times of
all four particle species. 
 Three parameters are required for this: the time offset, (defined
by the arrival time of a particle going at the speed of light), the length of
the beam-line and the beam momentum. 
The results are shown in
figure~\ref{fig:calcto4}, where a momentum of 168.9~MeV/c is found and the four
points are seen to be consistent. This run was taken with the final dipole
field reduced by 2\%, so under normal running conditions the beam momentum is 
172$\pm$2 MeV/c.

The spread in momentum can be estimated by noting that
it affects the velocity of pions but not electrons.
The quadrature difference in width of the pion and electron
peaks calculated as in figure~\ref{fig:tofmom} but after accounting for
small time-walk effects gives the spread in momentum as being between
1.0 and 1.2~MeV/c.


\subsection{Beam composition}
\label{sec:composition}

\begin{figure}[th]
\begin{center}
\epsfig{figure=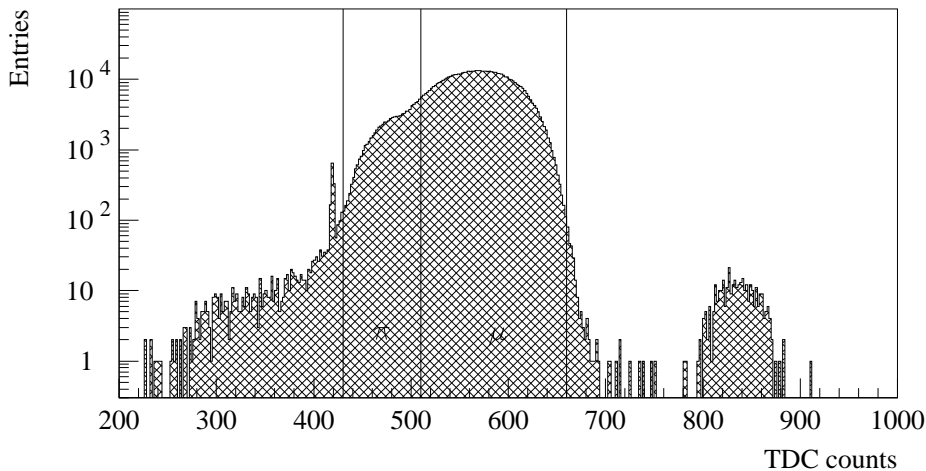,width=13.8cm}
\vspace*{-1cm}
\end{center}
\caption{\label{fig:evsel} The time of arrival of the beam cycle signal
relative to the \SoneB\ scintillator for run 290, an ordinary run
with no target. The graph
 corresponds to events where the digitised time from \SoneB\ was correct.
The section labelled $\mu$ contains those events used for the scattering
 analysis.
The small peak to the right contains electrons. The resolution is slightly
worse than in figure~\ref{fig:tofmom}; this is not understood.}
\end{figure}

The recorded sample under normal conditions has a large proportion of
 muons. This can be seen in
figure~\ref{fig:evsel}. Protons are ranged out in the \SoneA\  trigger,
and not generally recorded.
This figure shows a somewhat worse time resolution than in 
figure~\ref{fig:tofmom}, which is not understood but was constant
for the data used for the main analysis.


The muon fraction is further raised by demanding that the recorded time of the
\tof\ signal is within the section labelled $\mu$ in the figure.
The energy recorded in \Tina\ for the
subset of events within its acceptance is used 
to evaluate the purity of this sample. The left hand plot on 
 figure~\ref{fig:compos} (pion sideband), has
two large peaks  due to pions (left) and muons (right). The distribution is 
fitted using two Gaussian distributions to determine the peak positions and
widths. The relative amplitudes of these contributions are then fitted to
the right-hand plot, where a 0.8\%  pion component is found.
 Some contribution to the second
Gaussian comes from resolution tails, and therefore the pion contamination
is almost certainly overestimated.

\section{Analysis Method}
\label{sec:analysis}

\subsection{Event Selection}

The selection of events for analysis requires that they pass a number of 
selection criteria. Firstly the high voltage must have been nominal and the
data-acquisition in a normal state. Then we require that the event
was not taken during a period identified as having a problem with the time of
flight system. 

Next the number of reconstructed points on the first plane of the tracker
is required to be one or two; more hits would make the position of the
muon very ambiguous.
 Then the summed signal from the two
internal veto scintillators had to be  consistent with the pedestal,
and the time from \SoneB\ within a few counts  (0.2~ns) of nominal.

The hit with the largest light yield is taken as the estimated
particle position at each detector plane.
Finally, the positions of the hits on the first
and second planes are required to be consistent 
with the hypothesis of a particle coming from the target.
The second plane is about 4\% further from the target,
so it is required that  0.96 times the y position on the
second plane be within 5~mm of the y position on the first plane.
This procedure  is simpler than performing a full track fit, and 
less dependent upon alignment systematics.

\subsection{Deconvolution technique}

It is possible to compare the distributions of hits 
at the first detector plane.
However, the more fundamental quantity is the angle through which 
the particle scattered   in the target, and the effects of 
beam width, efficiency, resolution etc. must be corrected for 
to see this. This deconvolution is done using the \G4\ simulation, 
with the following approximate formula:

\begin{equation}
\overline{D} = \overline{B} + \overline{D_{\pi}} +  \overline{\overline{R}}
\cdot \overline{\overline{\epsilon}} \cdot \overline{\Theta}
\label{eq:docon}
\end{equation}

Where $\overline{D}$ is the observed position data, 
$\overline{B}$ a background of particles not passing through the target, 
$\overline{D_{\pi}}$ is the contamination from pions,
$\overline{\overline{R}}$ 
is the response of the detector to a particle deflected through angle
$\theta_y$, $\overline{\overline{\epsilon}}$ is the efficiency of the 
detector for particles deflected through angle $\theta_y$, and
$\overline{\Theta}$ is the projected scattering distribution in the 
target.

The background, $\overline{B}$, is found from simulation as
those muons which managed to meet the trigger conditions without passing
through the target. There are typically 0.125\% background events.
The pion contamination, $\overline{D_{\pi}}$, is taken from the data 
pion sideband. The default value is   0.8\% pions, and as the sideband
is approximately 50\% pions, a 1.6\% admixture of the data sideband is
added to
the simulation.

The response and efficiency matrices, $\overline{\overline{R}}$ 
and $\overline{\overline{\epsilon}}$ are taken from simulation.
The efficiency is found by running \G4\ twice. The first run 
is a simulation of the target only, illuminated by a monochromatic
collinear beam of muons. This is used to find the true $\overline{\Theta}$
distribution in
\G4. The second run is the full simulation, including trigger 
and tracking acceptance cuts. The distribution $\overline{\Theta}$
for those muons accepted is calculated, and the ratio of these is
$\overline{\overline{\epsilon}}$. In the above formula it is treated as a
diagonal matrix. 
Finally, $\overline{\overline{R}}$  is filled as a 2D matrix, 
 with each accepted simulated event being entered at a point 
given by its $\theta_y$ and the y position at the first measurement
plane. It therefore gives the probability of a given measured position
for each possible true deflection angle.

The equation is solved for $\overline{\Theta}$, imposing a requirement
of symmetry about $\theta=0$, using Minuit\cite{ref:Minuit},
which also finds errors and correlations. It has been checked that the
technique is mathematically correct, in that if given simulated data as an
input the scattering distribution from that simulation is recovered.

This formalism does not explicitly include  the  effect of energy losses and 
scattering in the $X$ direction. The matrices 
 $\overline{\overline{R}}$ and $\overline{\overline{\epsilon}}$ 
therefore depend on the simulation of multiple
scattering and energy loss in the \Geant\ model. In particular, the
probability that a particle completely misses the detector in the $X$
direction as a function of its $Y$ angle must be taken from simulation.

The problem
of instabilities in the deconvolution 
has been dealt with by reducing the number of bins in the
deconvoluted data, where 21 are used, compared to the raw data, which has 57.
Furthermore the condition of symmetry is enforced in the fit, so there are
essentially only 11 bins. Finally, the outermost bin in true scattering 
angle runs from 115~mrad to $\pi$, but is very weakly constrained
by the data as the majority of events in this bin are not recorded. It
also has a  strong dependence on the simulated prediction of the fraction
recorded.
It is not shown in the plots which follow.

\section{The results}
\label{sec:results}

\subsection{Comparisons at detector level}

\begin{figure}[htp]
\begin{center}
\epsfig{figure=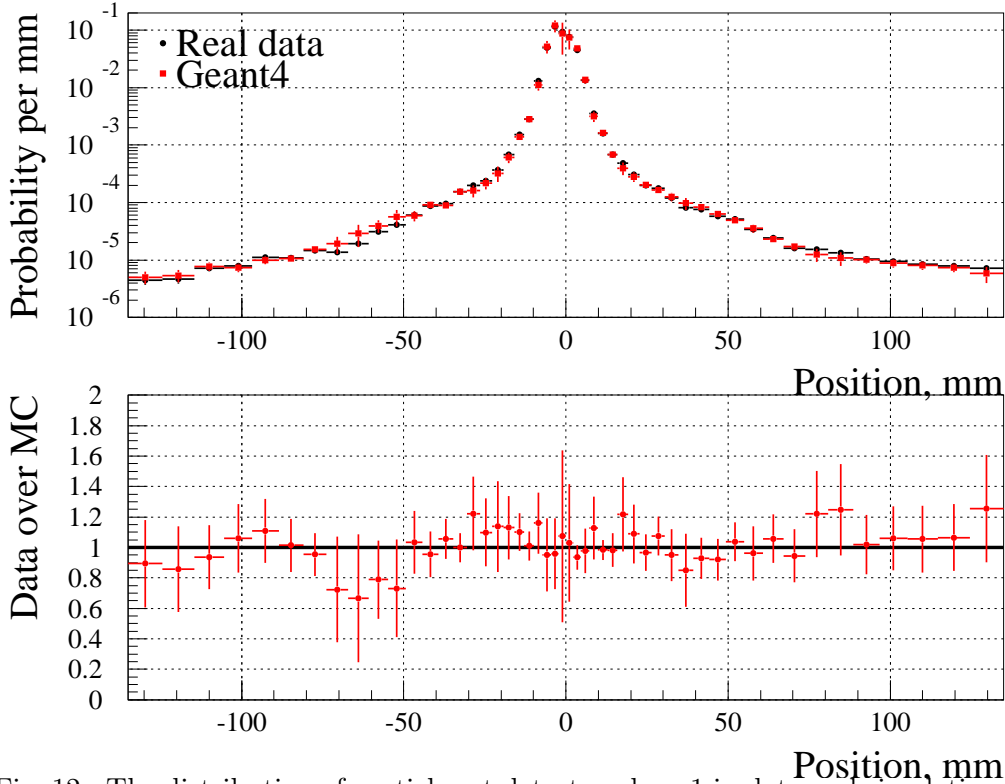,width=13.8cm}
\vspace*{-1cm}
\end{center}
\caption{\label{fig:targ_h2le} The distribution of particles at detector plane
1 in data and simulation for the empty H$_2$ vessel in the 150~mm position.
}
\end{figure}

Data taken with no target, an empty hydrogen vessel or a very thick iron
target are not of great intrinsic interest,
but were necessary to understand the detector. The first, the 
bare collimator distribution, is shown in figure~\ref{fig:bare_collimator}.
The final results are based upon deconvoluting the data using the simulation,
and so discrepancies here will translate into systematic errors on the results.
In the case of hydrogen data, the equivalent of the bare collimator 
is the empty target vessel. The distributions obtained for this
in data and simulation are showing in figure~\ref{fig:targ_h2le}
for the case of the longer target configuration. 

The error bars shown in these plots are statistical only for the data but
are a combined statistical and systematic error on the simulation points.
This systematic error explicitly includes the difference between these plots,
and so by construction all ratios must be compatible with one to within
one combined error.

The other control plot is
figure~\ref{fig:thick_steel}, which shows the
results for a thick iron target, of 28\%~X$_0$.
The agreement is
 well within 10\% in most bins, with the centre and the 
 edge bins  
showing differences around 20\%. 
As the edge bins have also got a large background subtraction, 
the region of the detector used for analysis and subsequent figures is
restricted to the central 270~mm.

\begin{figure}[htp]
\begin{center}
\epsfig{figure=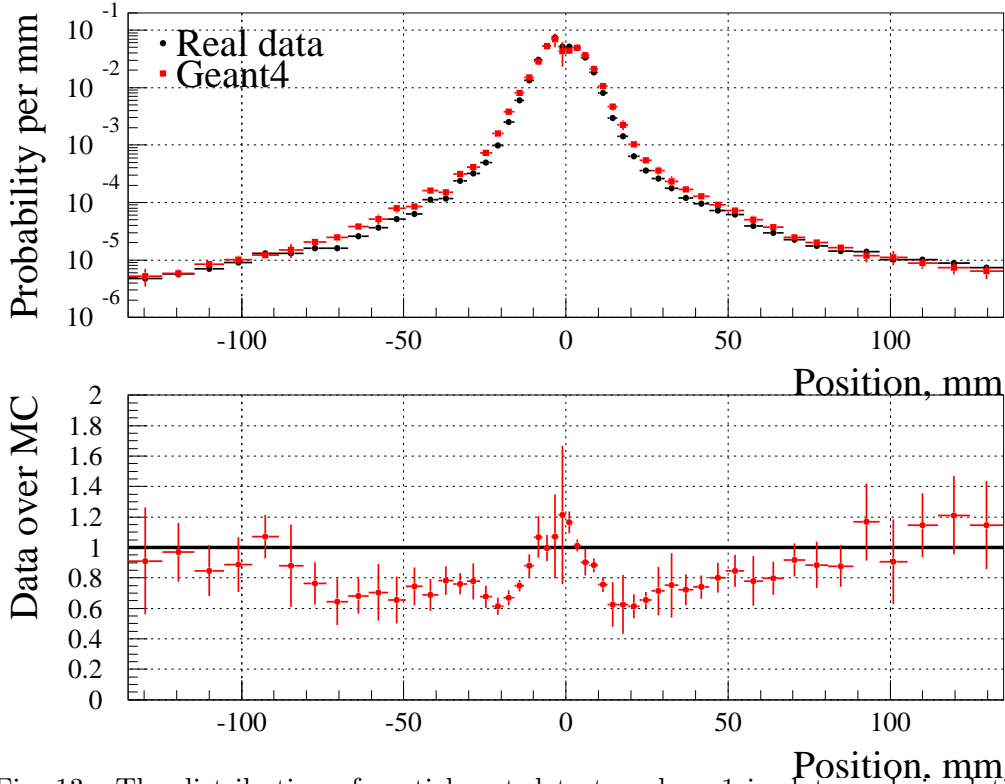,width=13.8cm}
\vspace*{-1cm}
\end{center}
\caption{\label{fig:targ_thin_lith} The distribution of particles at detector
plane 1 in data and simulation for thin lithium, combining both targets.
This distribution is used to check the agreement of our data with \Geant.
}
\end{figure}

The raw distribution at detector level for a few of the targets are now shown.
Figure~\ref{fig:targ_thin_lith} shows the combined data from the two 
 thin lithium targets.
The thickness of this sample corresponds to the second lowest fraction 
of a radiation length of material used, and it therefore
has a narrow distribution.
\Geant\ tends
to overestimate the distribution, starting from about 10~mrad.
Note that outside  about 100~mm the distribution is completely dominated by the
scattering from the collimator, see figure~\ref{fig:bare_collimator}, and the
 contribution from the lithium is
negligible. Thus the convergence of data and simulation in the tails 
is more related to the simulation of the collimator than the physics of
scattering.

\begin{figure}[htp]
\begin{center}
\epsfig{figure=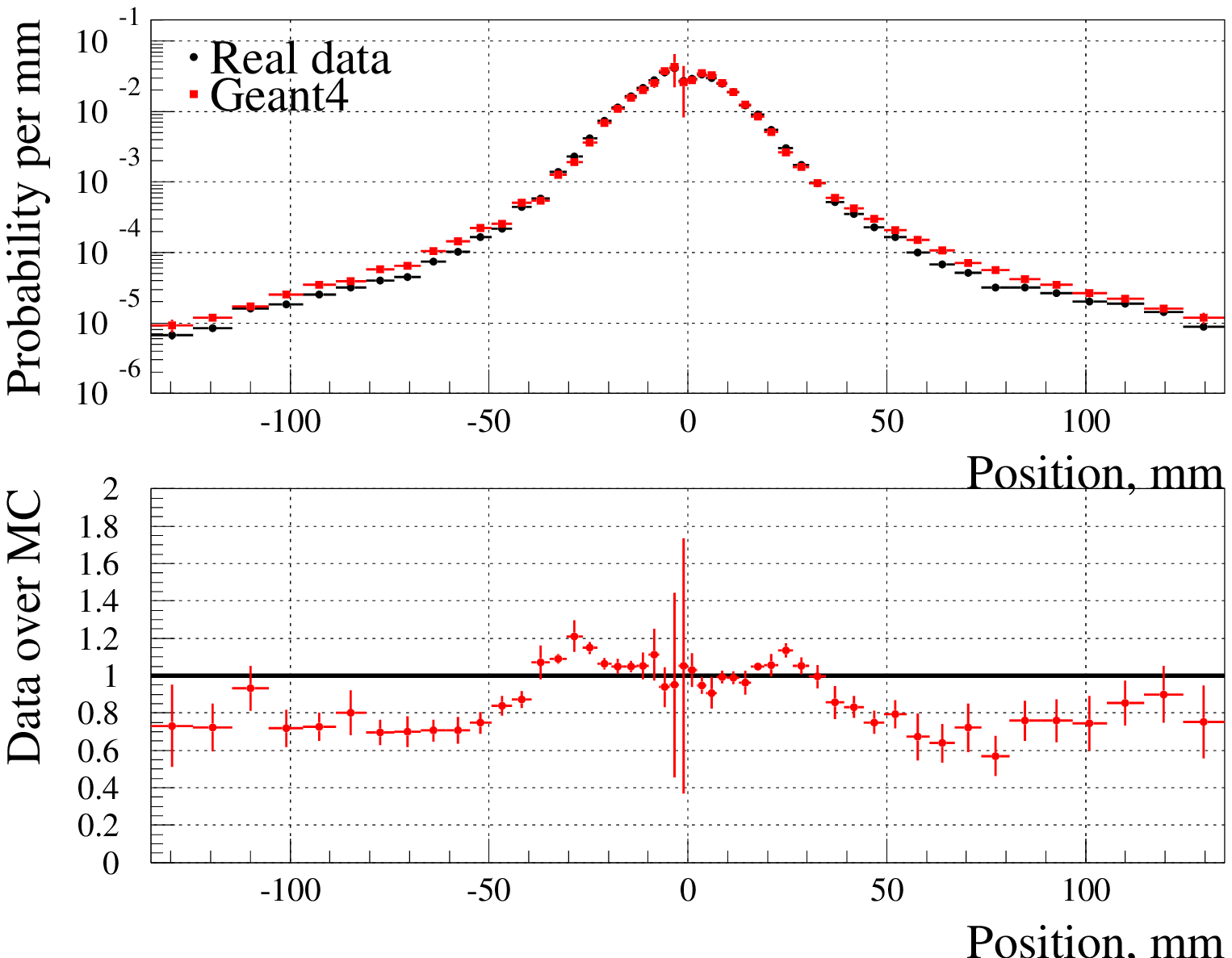,width=13.8cm}
\vspace*{-1cm}
\end{center}
\caption{\label{fig:targ10} The distribution of particles at detector plane
1 in data and simulation for thin Fe, target 10
}
\end{figure}

The thin iron target, figure~\ref{fig:targ10}, is the highest $Z$
target studied. The \Geant\ description is better than for any of the lighter
targets, but there is here a small region below 40~mm where the scattering 
distribution is underestimated which is seen in no other target.



\begin{figure}[htp]
\begin{center}
\epsfig{figure=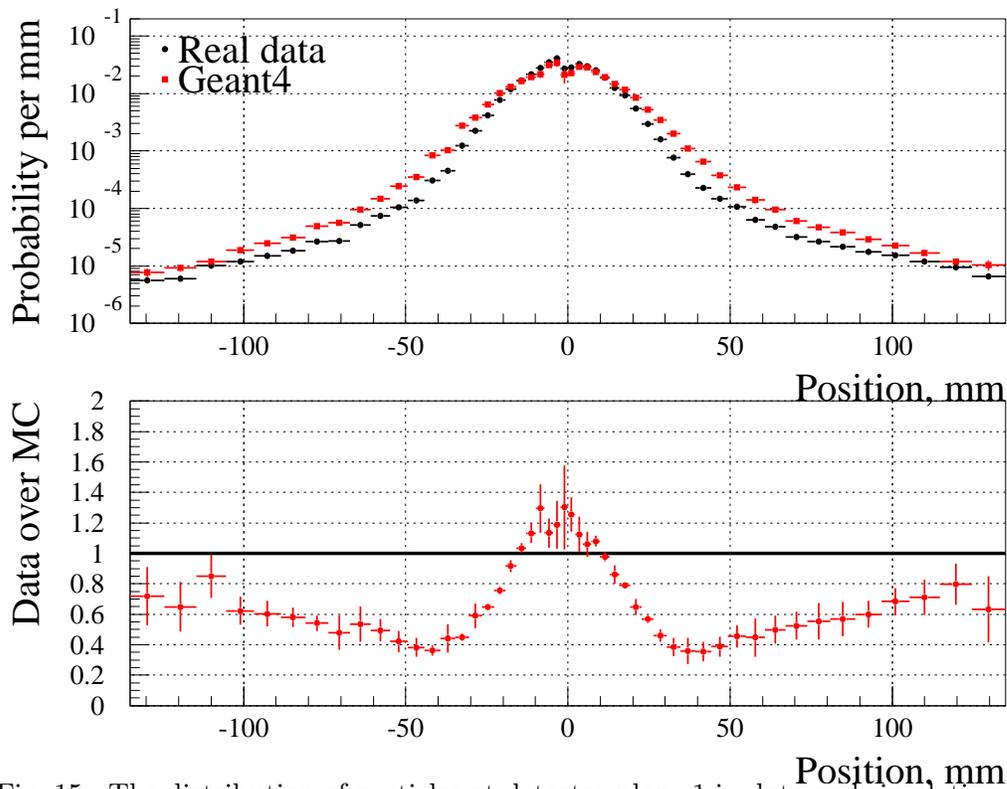,width=13.8cm}
\vspace*{-1cm}
\end{center}
\caption{\label{fig:targ_h2lf} The distribution of particles at detector plane 1 in data and simulation for 159~mm of liquid H$_2$.
}
\end{figure}

The final comparison presented at the level of the detector 
is for the longer hydrogen  target. 
This resembles the lithium target, but the differences between
data and simulation are  more
pronounced, with the observed distribution less than half the  predicted 
 level at $\pm$40~mm.

\subsection{Deconvoluted results}

\begin{figure}[htp]
\begin{center}
\epsfig{figure=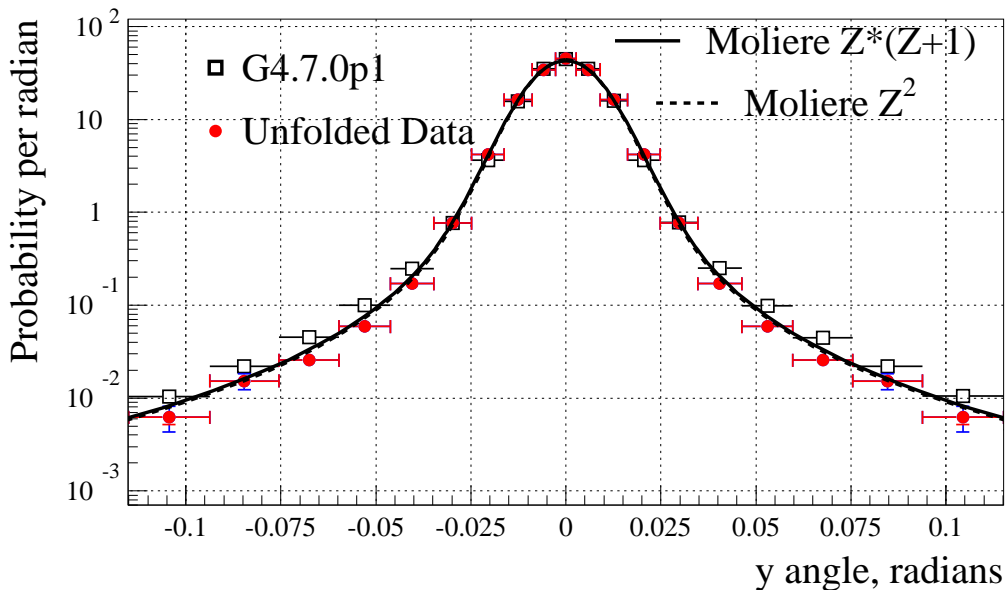,width=13.8cm}
\end{center}
\caption{\label{fig:unfold10} 
The projected scattering angle distribution 
 in data and simulation for thin iron, target 10.
}
\end{figure}

The deconvolution procedure described in section~\ref{sec:analysis}
is applied to each of the targets. A series of systematic effects are allowed
for, and plots for many of the targets 
are presented in figures~\ref{fig:unfold10}
to~\ref{fig:unfold_h2lf}. In each case the statistical error is indicated with
an inner error bar, and the combined statistical plus systematic error is
indicated with the outer error bar.
In most cases the statistical errors are only important for regions
of the distribution where the probability of arrival is  below
about 0.1 per radian.
The data for all targets are shown in table~\ref{ta:results3}.


The plots have two smooth curves superimposed: the solid line represents the
Moliere calculation 
with a  $Z\times(Z+1)$ factor for the large angle scatters,
as proposed by Bethe\cite{ref:bethe},
and the dashed line represents the same calculation using a $Z^2$ term.
The difference corresponds to scattering solely on the nucleus, or
also on the electrons.
Note that the kinematics of scattering prevent a muon from making
a single scatter from a free electron through an angle greater than
$m_e/m_\mu$, or 5~mrad. We therefore expect that for the description of the
tail region, above 15 or 20~mrad, the  $Z^2$ term will be more appropriate.

\newcounter{beanN}
\begin{list}
{\Roman{beanN}}{\usecounter{beanN}
\setlength{\rightmargin}{\leftmargin}}

\item{\bf Thin Iron}. The data are shown in figure~\ref{fig:unfold10}. 
The two Moliere models are almost indistinguishable, and
provide a good description of the data. In general, the \Geant~4.7.0p1
 simulations
are somewhat in excess of the data above 35~mrad.
Errors are around 5\%, growing to 30\% in the final bin shown.

\begin{figure}[htp]
\begin{center}
\epsfig{figure=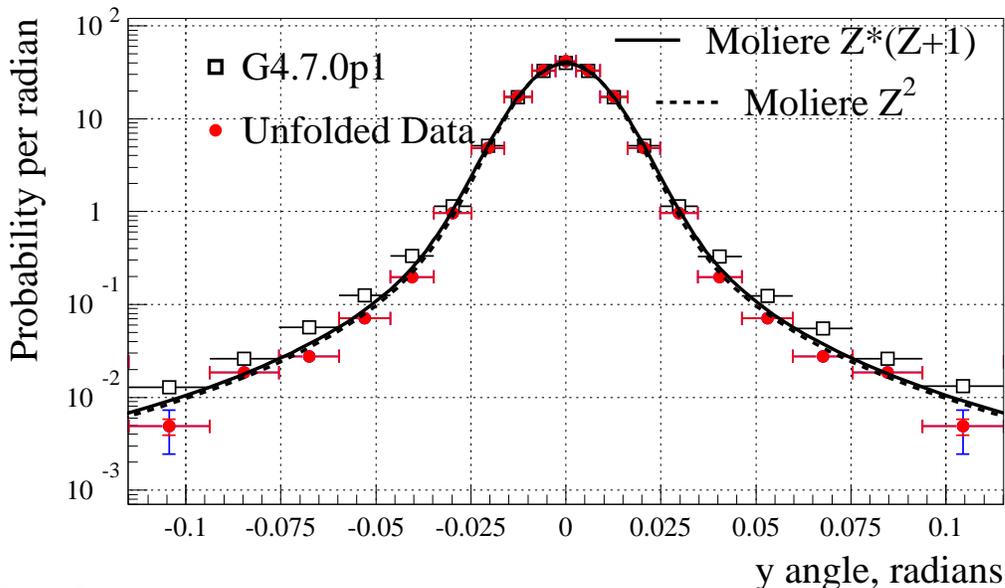,width=13.8cm}
\vspace*{-1cm}
\end{center}
\caption{\label{fig:unfold8} 
The projected scattering angle distribution 
 in data and simulation for aluminium, target 8.
}
\end{figure}

\item{\bf Aluminium}. The  data, figure~\ref{fig:unfold8}, 
are rather similar to the iron, with an overestimation
of the tail by \G4\ of around a factor two.



\begin{figure}[htp]
\begin{center}
\epsfig{figure=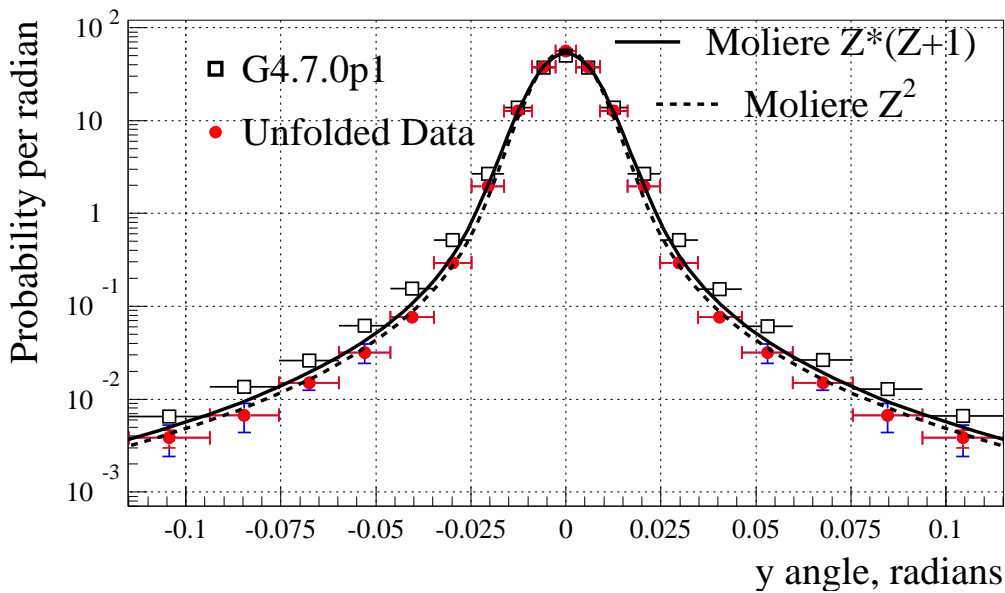,width=13.8cm}
\end{center}
\caption{\label{fig:unfold7} 
The projected scattering angle distribution 
 in data and simulation for carbon, target 7.
}
\end{figure}

\item{\bf Carbon}. The data, shown in figure~\ref{fig:unfold7},  reveal a 
visually apparent distinction between the Moliere models.
Furthermore, the
\Geant\ simulation  appears to have an excess compared with 
data starting at  lower angles, around 20~mrad, than it did for iron.


\item{\bf Polyethylene} 
The data for this target resemble beryllium, with a similar 
overestimate from \Geant\ of the scattering. They are not shown.
The largest systematic contributions come from the simulated
collimator description.

\begin{figure}[htp]
\begin{center}
\epsfig{figure=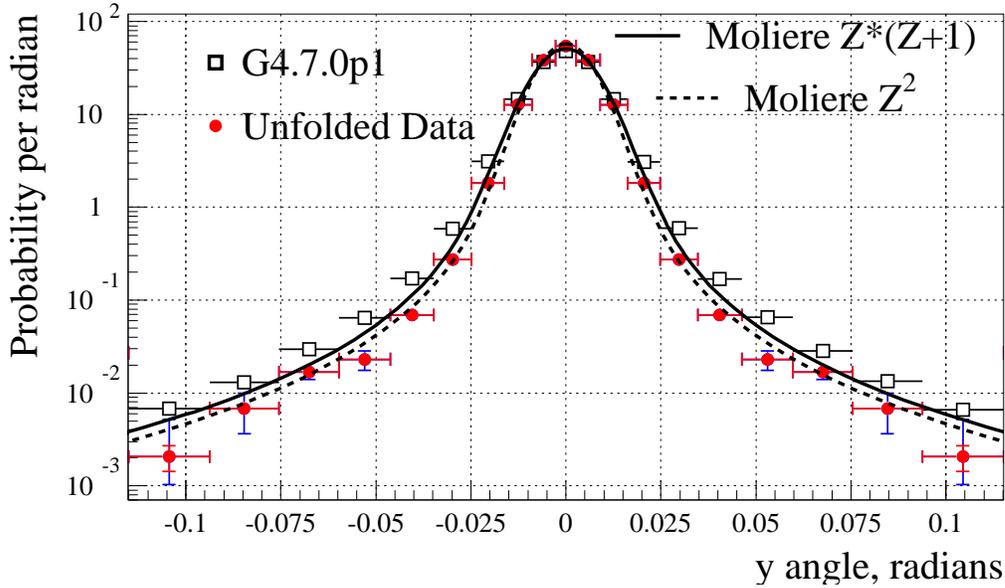,width=13.8cm}
\end{center}
\caption{\label{fig:unfold5} 
The projected scattering angle distribution 
 in data and simulation for thick beryllium, target 5.
}
\end{figure}

\item{\bf Beryllium}
The thick target data are shown in 
figure~\ref{fig:unfold5}. 
The \Geant\ model overstates the tails by approximately a factor of two.
Of the two Moliere models, the  $Z^2$ is slightly  preferred.
No distinction was possible for materials with higher $Z$
as the models were  similar.
The thin beryllium has  larger errors, owing to the large background 
subtraction required there, and is not displayed.

\begin{figure}[htp]
\begin{center}
\epsfig{figure=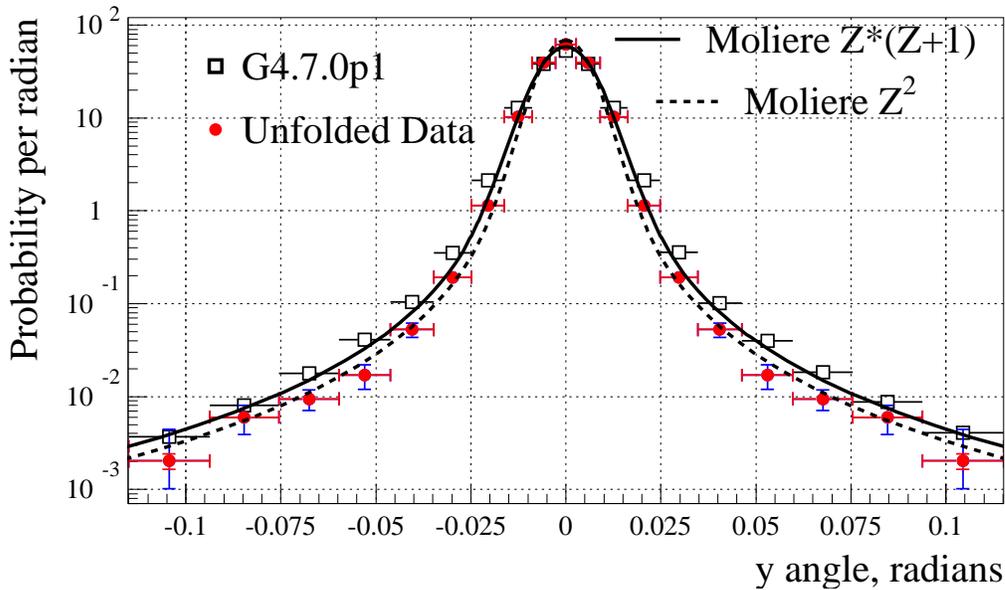,width=13.8cm}
\end{center}
\caption{\label{fig:unfold3} 
The projected scattering angle distribution 
 in data and simulation for thick lithium, both targets combined.
}
\end{figure}

\item{\bf Lithium} There were two pairs of lithium targets with 
very similar
thicknesses. The data for the thicker pair are combined in
figure~\ref{fig:unfold3}.
The Moliere model with a $Z^2$ term provides a reasonable
description of the data,
but  the $Z\times(Z+1)$ Moliere model is now clearly disfavoured.
\Geant\ continues to have tails about a factor two above the data.
The data from the thin lithium targets have large systematic and
statistical errors
in the tails, owing to the large fraction of background which has been
accounted for in the deconvolution, and  are not shown.

\begin{figure}[htp]
\begin{center}
\epsfig{figure=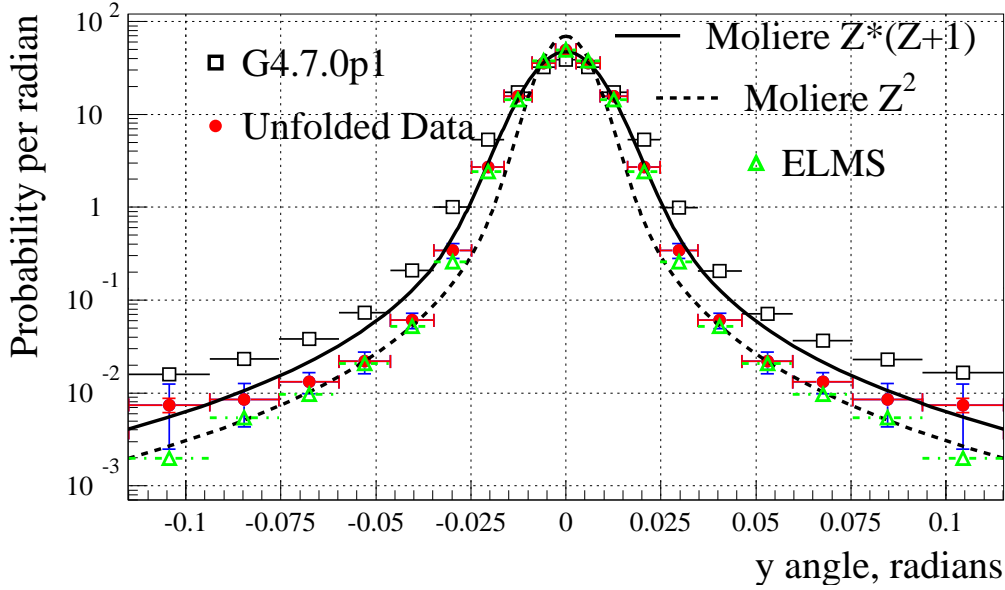,width=13.8cm}
\end{center}
\caption{\label{fig:unfold_h2sf} 
The projected scattering angle distribution 
 in data and simulation for 109~mm of liquid H$_2$.
}
\end{figure}

\item{\bf Hydrogen} For 109~mm of hydrogen it can be clearly seen
in figure~\ref{fig:unfold_h2sf} that neither
Moliere model accurately describes the data. The $Z^2$ model understates the
 region below   around 25~mrad, and above 60~mrad, although here the 
errors are much larger.
\Geant~4.7.0p01
predicts too much scattering, by a factor of around 4
at 50~mrad.

The data from the ELMS model\cite{ref:elms} are also compared with our
results. These seem to show good agreement at all but the 
high angle. This region is has the largest systematic errors.

\begin{figure}[htp]
\begin{center}
\epsfig{figure=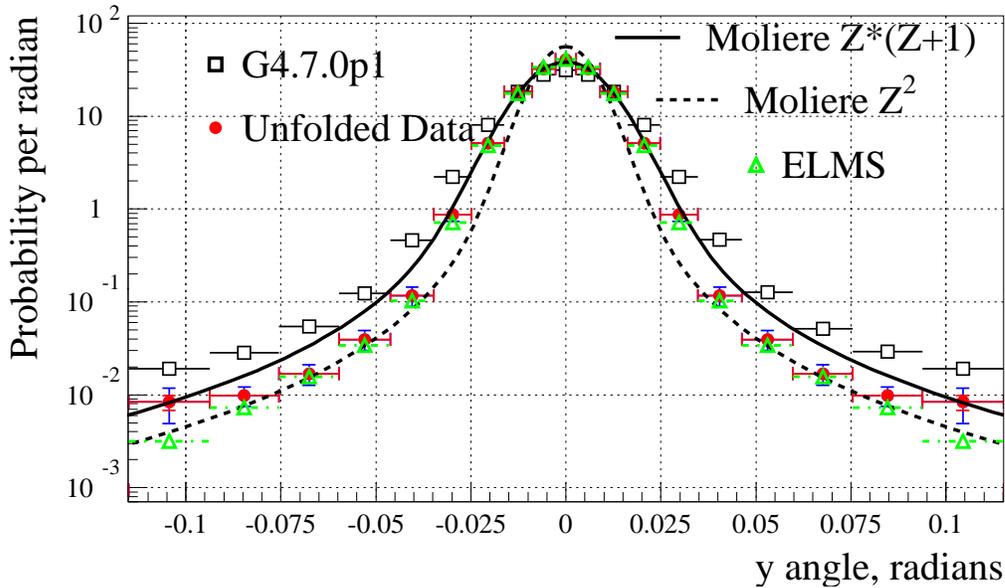,width=13.8cm}
\end{center}
\caption{\label{fig:unfold_h2lf} 
The projected scattering angle distribution 
 in data and simulation for 159~mm of liquid H$_2$.
}
\end{figure}

It can be clearly seen in figure~\ref{fig:unfold_h2lf} that neither
Moliere model accurately describes the data for the long hydrogen target.
 The $Z^2$ model seems
correct beyond around 40~mrad, but in the multiple-scattering dominated
region the $Z\times(Z+1)$ model is much more appropriate. \Geant~4.7.0p01
predicts too much scattering, but close examination reveals that the excess is
less than for the thinner sample.

\end{list}

Overall, we see that when the two Moliere models are distinguishable,
i.e. when $Z$ is small,
the data is described by the $Z\times(Z+1)$ in the Gaussian core, while
the tails show better agreement with the $Z^2$ version. This emphasises that
the difficulty of describing the data is greater at low $Z$.

\Geant~4.7.0p01 typically predicts about a factor two more scattering
tail than is observed. This is not a very large  effect  given that the 
data cover 4 orders of magnitude. However, in the case of liquid hydrogen
the discrepancy grows to be a factor 4. 

The ELMS simulation describes hydrogen very well.

\subsection{Systematic Errors}

Systematic errors are evaluated by performing the complete analysis 
with different treatment of the data. The difference with the
main results is taken as an estimate of systematic error. All
errors are combined in quadrature.

\subsubsection{Collimator Description}

Three different techniques are used to evaluate the impact 
of differences between the collimated distributions seen in
figure~\ref{fig:bare_collimator}. The largest systematic errors
are generally from uncertainty in our understanding of the collimated beam.

\newcounter{bean}
\begin{list}
{A--\Roman{bean}}{\usecounter{bean}
\setlength{\rightmargin}{\leftmargin}}


\item{\bf Collimator difference}
The bare collimator distribution seen in simulation is subtracted from the
observed data. This residual is then added to the simulated distribution
for whichever target is currently under study. This is not completely correct,
as it ignores the scattering in
the target, but it provides an  estimate of the systematic effects.

\item{\bf Collimator deconvolution}
The deconvolution to find the underlying scattering distribution
is performed on the data with no target present. This would yield
a delta function, but does not due to bin width effects and discrepancies
between data and simulation.
The same is done in
simulation, and the difference of the two is found. This difference
is due to poor modelling of the collimated beam.
The deconvolution
for any given target is then repeated with this difference distribution
included as an extra convolution term on the right hand side of
equation~\ref{eq:docon}. Any changes in the results are taken
as systematic errors.

\item{\bf Internal veto scintillators}
The collimation system, described in section~\ref{sec:coll},
 has two internal veto scintillators mounted
just before the second collimator. The data show a good separation 
between pedestal and hits, but there is no complete cross-check on this.
The simulated efficiency is lowered to 90\%, (which  also provides a
slightly better description of the scattering distributions) and a systematic
error is generated.

\end{list}

\subsubsection{Detector alignment}

Detector misalignment relative to the simulation will cause errors in the
deconvolution.

\newcounter{beanB}
\begin{list}
{B--\Roman{beanB}}{\usecounter{beanB}
\setlength{\rightmargin}{\leftmargin}}

\item{\bf Detector X alignment}
Any credible X misalignment will have negligible impact.

\item{\bf Detector Y alignment}
A transverse shift of the detector of 0.1~mm is allowed as a systematic
error. 
The size comes from inspection of figure~\ref{fig:bare_collimator}.

\item{\bf Detector Z alignment}
The Z position of the detector affects the widths of the distribution.
From a combination of studying internal distributions in the tracker
and knowledge of the detector construction a systematic error of 5~mm
has been assigned.

\end{list}

\subsubsection{Detector modelling}

\newcounter{beanC}
\begin{list}
{C--\Roman{beanC}}{\usecounter{beanC}
\setlength{\rightmargin}{\leftmargin}}

\item{\bf Thick target comparison}
The muon tracking efficiency 
maps directly into the scattering distribution, and as such it must be well
controlled. The distributions in data and simulation observed with the 
thick iron target present, figure~\ref{fig:thick_steel}, are used to 
monitor this. It is conservatively assumed that all differences are
due to efficiency, and so the ratio of these distributions
is used as an efficiency correction. The change in the unfolded 
results when this is applied is regarded as a  systematic 
error.

\item{\bf Cross-talk suppression}
The  default cut selection allows one or two hits to be reconstructed on the
first detector plane. When we require exactly one the probability of
picking the wrong one is reduced. The analysis is repeated with this
requirement and the difference is taken as a systematic error.



\end{list}

\subsection{Pion contamination}
As discussed in section~\ref{sec:composition}, a pion background of
0.8\% is estimated. However, this comes from the
observation that figure~\ref{fig:compos} has tails to the left
of the muon peak; some of this background will arise from sources other than
 pions.
The most extreme alternative assumption is that there are no pions present,
and this is used to estimate a systematic error.

\subsection{Deconvolution matrix}
The matrix used to relate true scattering angle to measured position
is taken from simulation. It depends   upon the physics in
that simulation, especially  scattering in the  X direction.
This dependence is reduced by re-weighting all the simulated events
as a function of the  space angle scatter
with an ad-hoc simple function  which moves \G4\ towards
Moliere $Z^2$ scattering above about 40~mrad.
 The central values of all the fits have been
obtained using this procedure. A systematic error 
is assigned by comparing with 
the unweighted \Geant, except in the case of hydrogen where we have 
the ELMS simulation, and the alternative distribution is \G4\ re-weighted to
match the predictions of that.
This is typically the largest systematic error at the edge of the deconvoluted
distributions.

\section{Conclusions}
\label{sec:conclusions}

These data allow the validation of codes to calculate multiple scattering in
the regime relevant for ionisation cooling. In general Moliere scattering
provides a good description, but for $Z$ below about 4 there are differences 
which may be important. In particular, for hydrogen neither Moliere model used
is appropriate. 
The ELMS simulation provides a  good prediction of  the deconvoluted hydrogen
distributions.

The data show noticeable deviations from the predictions of the \Geant~4.7.0p01
simulation code, which tends to overstate the scattering tail by about
a factor of two, and more strongly
the lower the $Z$ of the material. For hydrogen it is about a factor of 4.

The discrepancies observed previously\cite{ref:andrievsky} are not confirmed.
Indeed, the scattering predicted by the \Geant\ simulation needs to be
reduced to match the data, not increased.

\section{Acknowledgements}

We would like to thank all the staff of TRIUMF Laboratory for the helpful
and friendly way in which we were welcomed to the laboratory and for
all the facilities that
were put at our disposal.
 The following all lent support, for which we are most grateful: 
M-P.Boudet,  R.Barlow, S.Burge, J.Flynn, 
R.Hampton, D.Laihem, S.Malton, C.Marshall, E.O'Neill, P.Rock,
W.Sievers, D.Wade, P.Zubko

We also acknowledge the financial support from PPARC and CCLRC.


\section{Appendix }

\begin{table}[htbp]
\begin{center}
\tiny
\begin{tabular}{rrrcccccc}     \hline
 &    X$_0$ &     & \multicolumn{6}{c}{Upper edge of bin, radians} \\
 & \%       & Type  & 0.00269 & 0.00895 & 0.0162 & 0.0248 & 0.0347 & 0.0463 \\
\hline \hline 
\multirow{4}{*}{Li}
& \multirow{2}{*}{0.41}
 & data &89.4$\pm$4.4&37.3$\pm$1.7&3.49$\pm$0.27&0.31$\pm$0.05
&0.08$\pm$0.02&0.017$\pm$0.009\\
 &  & G4  &75.6$\pm$0.1&39.7$\pm$.04&5.64$\pm$0.01&0.56$\pm$.004
&0.13$\pm$.002&0.046$\pm$0.001\\
& \multirow{2}{*}{0.82}
   & data &61.7$\pm$3.4&39.3$\pm$1.3&10.3$\pm$0.21&1.14$\pm$0.06
&0.19$\pm$0.02&0.053$\pm$0.009\\
 &  & G4   &53.1$\pm$.04&38.3$\pm$.03&12.9$\pm$0.02&2.13$\pm$.007
&0.36$\pm$.003&0.101$\pm$0.001\\
\hline
\multirow{4}{*}{Be}
& \multirow{2}{*}{0.28}
 & data &112.$\pm$4.9&29.9$\pm$2.0&1.43$\pm$0.17&0.16$\pm$0.04
&0.04$\pm$0.014&0.013$\pm$0.011\\
 &  & G4  &95.1.$\pm$0.1&35.4$\pm$.03&2.55$\pm$0.01&0.33$\pm$.003
&0.09$\pm$0.002&0.033$\pm$0.001\\
& \multirow{2}{*}{1.06}
  & Data & 54.2$\pm$5.4&38.4$\pm$1.2&12.7$\pm$0.63&
1.82$\pm$0.17&0.27$\pm$0.025&0.069$\pm$0.010\\
& & G4 &48.0$\pm$0.05&36.5$\pm$0.04&14.6$\pm$0.02&
 3.10$\pm$0.01&0.60$\pm$0.004&0.170$\pm$0.002\\
\hline
\multirow{2}{*}{CH$_2$} 
 & \multirow{2}{*}{0.99} 
 & data &55.4$\pm$3.1&38.2$\pm$1.3&12.6$\pm$0.18&1.79$\pm$0.10
&0.28$\pm$0.020&0.073$\pm$0.010\\
 & & G4 &50.2$\pm$.05&37.2$\pm$.04&13.9$\pm$0.02&2.68$\pm$.009
&0.49$\pm$0.004&0.140$\pm$0.002\\
\hline
\multirow{2}{*}{C} 
 & \multirow{2}{*}{1.53} 
 & data &55.7$\pm$3.0&37.8$\pm$1.3&12.7$\pm$0.18&1.97$\pm$0.08
&0.29$\pm$0.022&0.076$\pm$0.010\\
 & & G4 &50.3$\pm$.05&37.2$\pm$.04&13.7$\pm$0.02&2.67$\pm$.009
&0.52$\pm$0.004&0.153$\pm$0.002\\
\hline
\multirow{2}{*}{Al} 
 & \multirow{2}{*}{1.69} 
 & data &41.7$\pm$2.4&33.0$\pm$1.0&17.2$\pm$0.20&4.82$\pm$0.10
&0.96$\pm$0.040&0.198$\pm$0.014\\
 & & G4 &39.8$\pm$.04&32.9$\pm$.03&17.0$\pm$0.02&5.14$\pm$0.01
&1.15$\pm$0.006&0.327$\pm$0.003\\
\hline
\multirow{2}{*}{Fe} 
 & \multirow{2}{*}{0.82} 
 & data &45.4$\pm$2.8&34.4$\pm$1.1&16.3$\pm$0.15&4.19$\pm$0.10
&0.76$\pm$0.034&0.171$\pm$0.012\\
&  & G4 &44.4$\pm$.05&35.3$\pm$.04&15.8$\pm$0.02&3.64$\pm$0.01
&0.78$\pm$0.004&0.250$\pm$0.002\\
\hline
\multirow{6}{*}{H$_2$}
& \multirow{3}{*}{1.31}
 & data &49.5$\pm$2.7&35.8$\pm$1.5&15.7$\pm$0.41&2.70$\pm$0.39
&0.34$\pm$0.061&0.061$\pm$0.012\\
& & G4 &38.7$\pm$.04&32.2$\pm$.03&17.4$\pm$0.02&5.33$\pm$0.01
&1.00$\pm$0.005&0.205$\pm$0.002\\
& & ELMS &49.6$\pm$.07&37.7$\pm$.05&14.5$\pm$0.03&2.43$\pm$0.01
&0.26$\pm$0.004&0.052$\pm$0.001\\
& \multirow{3}{*}{1.90}
& data &40.6$\pm$1.7&32.3$\pm$0.9&18.3$\pm$0.18&5.10$\pm$0.38
&0.88$\pm$0.014&0.117$\pm$0.028\\
 & & G4 &31.4$\pm$.04&27.9$\pm$.03&18.3$\pm$.025&8.06$\pm$0.02
&2.22$\pm$0.008&0.466$\pm$0.003\\
& & ELMS &40.9$\pm$.06&33.8$\pm$.05&17.5$\pm$0.03&4.86$\pm$0.02
&0.72$\pm$0.006&0.103$\pm$0.002\\

\hline
 \hline
 &    X$_0$ &     & \multicolumn{5}{c}{Upper edge of bin, radians} \\
 & \%       & Type  & 0.0597 & 0.0754 & 0.0938 & 0.1151 & 3.141 \\
\hline \hline 
\multirow{4}{*}{Li}
& \multirow{2}{*}{0.41}
 & data &
0.010$\pm$0.006&0.008$\pm$0.004
&0.004$\pm$0.003&0.003$\pm$0.002 &0.0026$\pm$0.0026\\
 &  & G4  &
0.018$\pm$0.001&0.008$\pm$.0004
&0.004$\pm$.0002&0.002$\pm$.0001 &1.8E-05$\pm$1.8E-06\\
& \multirow{2}{*}{0.82}
   & data &
0.017$\pm$0.005&0.009$\pm$0.002
&0.006$\pm$0.002&0.002$\pm$0.002 &0.0007$\pm$0.0007\\
 &  & G4   &
0.040$\pm$.0007&0.018$\pm$.0005
&0.009$\pm$.0003&0.004$\pm$0.002 &3.6E-05$\pm$1.8E-06\\
\hline
\multirow{4}{*}{Be}
& \multirow{2}{*}{0.28}
 & data &
0.006$\pm$0.005&0.001$\pm$0.003
&0.004$\pm$0.004&0.003$\pm$0.003 &0.00017$\pm$0.00017\\
 &  & G4  &
0.014$\pm$0.001&0.007$\pm$.0003
&0.003$\pm$.0002&0.002$\pm$.0001&1.8E-05$\pm$1.8E-06 \\
& \multirow{2}{*}{1.06}
  & Data &
0.023$\pm$0.006
&0.017$\pm$0.003&0.007$\pm$0.003&0.002$\pm$0.0031&0.0024$\pm$0.0024 \\
& & G4 &
0.065$\pm$0.001
&0.029$\pm$.0007&0.013$\pm$.0004&0.007$\pm$0.0003&7.2E-05$\pm$0.2E-05\\
\hline
\multirow{2}{*}{CH$_2$} 
 & \multirow{2}{*}{0.99} 
 & data &
0.023$\pm$0.005&0.013$\pm$0.002
&0.008$\pm$0.003&0.002$\pm$0.002 &0.0005$\pm$0.0005\\
 & & G4 &
0.054$\pm$0.001&0.025$\pm$.0006
&0.012$\pm$.0004&0.006$\pm$.0003 &5.4E-05$\pm$1.8E-06\\
\hline
\multirow{2}{*}{C} 
 & \multirow{2}{*}{1.53} 
 & data &
0.032$\pm$0.008&0.015$\pm$0.003
&0.007$\pm$0.002&0.004$\pm$0.001 &0.0011$\pm$0.0011\\
 & & G4 &
0.061$\pm$0.001&0.026$\pm$.0006
&0.013$\pm$.0004&0.007$\pm$.0003 &5.4E-05$\pm$1.8E-06\\
\hline
\multirow{2}{*}{Al} 
 & \multirow{2}{*}{1.69} 
 & data &
0.071$\pm$0.009&0.028$\pm$0.003
&0.018$\pm$0.003&0.005$\pm$0.002 &0.0008$\pm$0.0008\\
 & & G4 &
0.124$\pm$0.002&0.055$\pm$0.001
&0.026$\pm$.0006&0.013$\pm$.0004 &0.00010$\pm$3.6E-06\\
\hline
\multirow{2}{*}{Fe} 
 & \multirow{2}{*}{0.82} 
 & data &
0.059$\pm$0.008&0.026$\pm$0.003
&0.015$\pm$0.003&0.006$\pm$0.002&0.0014$\pm$0.0013 \\
&  & G4 &0.099$\pm$0.001&0.045$\pm$.0008
&0.022$\pm$.0005&0.011$\pm$.0004&9.0E-05$\pm$1.8E-06 \\
\hline
\multirow{6}{*}{H$_2$}
& \multirow{3}{*}{1.31}
 & data &
0.022$\pm$0.006&0.013$\pm$0.003
&0.008$\pm$0.004&0.007$\pm$0.005& 0.0025$\pm$0.0025\\
& & G4 &
0.071$\pm$0.001&0.036$\pm$.0008
&0.023$\pm$.0006&0.017$\pm$.0004&0.0024$\pm$1.4E-05 \\
& & ELMS &
0.021$\pm$.0009&0.010$\pm$.0006
&0.005$\pm$.0004&0.002$\pm$.0002& 1.8E-05$\pm$1.8E-06\\
& \multirow{3}{*}{1.90}
& data &
0.039$\pm$0.009&0.017$\pm$0.004
&0.010$\pm$0.002&0.008$\pm$0.003 &0.004$\pm$0.004\\
 & & G4 &
0.128$\pm$.0015&0.052$\pm$.0009
&0.029$\pm$.0006&0.019$\pm$.0005 &0.0025$\pm$1.5E-05\\
& & ELMS &
0.034$\pm$0.001&0.016$\pm$.0007
&0.007$\pm$.0004&0.003$\pm$.0003&3.6E-05$\pm$0.2E-05 \\

\hline
\end{tabular}
\caption[]{
The data and simulated predictions for the probability per radian
of scattering through a given projected angle. 
The table is split on two halves to fit the page.
The first bin starts
at zero. The errors on each bin are total errors. Note that the
last bin is essentially an overflow bin and has 100\% errors.
}
\label{ta:results3}
\end{center}
\end{table}


\newpage
\begin{thebibliography}{99}

\bibitem{ref:cooling}
D. V. Neuffer, Muon cooling and applications, Proc. Workshop
on Beam Cooling, Montreux 1993, CERN Rep. 94-03, 1994, p 49.

\bibitem{ref:andrievsky}
A. I. Andrievsky et al., J. Phys. (USSR) 6 (1942) 279

\bibitem{ref:moliere}
V. G. Moliere, Z. Naturforschg 3a (1948) 78; W. T. Scott, Rev. Mod. Phys. 35 (1963) 231

\bibitem{ref:fernow}
US Muon Collaboration technical note 123, 1998.
http://www-mucool.fnal.gov/notes/notes.html

\bibitem{ref:g4}
S. Agostinelli et al.,
Nucl. Instrum. Meth. {\bf A} 506 (2003) 250-303.


\bibitem{ref:elms}
W. W. M. Allison, J. Phys G; Nucl. Part. Phys. {\bf 29} (2003) 1701-1703. \\
and 
{\em http://www-pnp.physics.ox.ac.uk/~holmess/ELMS/}

\bibitem{ref:tina}
C. E. Waltham et al., 
Nucl. Instrum. Meth. {\bf A} 256 (1987) 91-97


\bibitem{ref:PMT}
Hammamatsu photonics
R5900U-L16 {\em Multianode 16 Channel Linear Array. }


\bibitem{ref:Minuit}
Minuit
{\em CERN Program Library entry {\bf D506}}

\bibitem{ref:bethe}
H. Bethe, 
Phys. Rev. 89, (1953) 1256-1266


\end{thebibliography}
\end{document}